\documentclass[iop,numberedappendix]{emulateapj}

\shorttitle{CO Survey of Local LIRGs}
\shortauthors{Yamashita et al.}
\begin{document}

\title{Cold Molecular Gas Along the Merger Sequence in Local Luminous Infrared Galaxies}

\author{Takuji Yamashita\altaffilmark{1,2}}\email{takuji@ir.isas.jaxa.jp}
\author{Shinya Komugi\altaffilmark{3}}
\author{Hideo Matsuhara\altaffilmark{1,4}}
\author{Lee Armus\altaffilmark{5}}
\author{Hanae Inami\altaffilmark{6}}
\author{Junko Ueda\altaffilmark{7}}
\author{Daisuke Iono\altaffilmark{8,9}}
\author{Kotaro Kohno\altaffilmark{10}}
\author{Aaron S. Evans\altaffilmark{11,12}}
\author{Ko Arimatsu\altaffilmark{8}}

\altaffiltext{1}{Institute of Space and Astronautical Science, Japan Aerospace Exploration Agency, 3-31-1 Yoshinodai, Chuo-ku, Sagamihara, Kanagawa 252-5210, Japan.}
\altaffiltext{2}{Research Center for Space and Cosmic Evolution, Ehime University, Bunkyo-cho, Matsuyama, Ehime 790-8577, Japan}
\altaffiltext{3}{Kogakuin University, 2665-1 Nakanocho, Hachioji, Tokyo 192-0015, Japan}
\altaffiltext{4}{Department of Physics, Tokyo Institute of Technology, 2-12-1 Ookayama, Meguro, Tokyo 152-8550, Japan}
\altaffiltext{5}{Spitzer Science Center, California Institute of Technology, Pasadena, CA 91125, USA}
\altaffiltext{6}{Centre de Recherche Astrophysique de Lyon (CRAL), Observatoire de Lyon, CNRS, UMR5574, F-69230, Saint-Genis-Laval, France}
\altaffiltext{7}{Harvard-Smithsonian Center for Astrophysics, 60 Garden Street, Cambridge, MA 02138, USA}
\altaffiltext{8}{National Astronomical Observatory of Japan, 2-21-1 Osawa, Mitaka, Tokyo 181-8588, Japan}
\altaffiltext{9}{The Graduate University for Advanced Studies (SOKENDAI), 2-21-1 Osawa, Mitaka, Tokyo 181-0015, Japan}
\altaffiltext{10}{Institute of Astronomy, The University of Tokyo, Osawa 2-21-1, Mitaka, Tokyo 181-0015, Japan}
\altaffiltext{11}{Department of Astronomy, University of Virginia, P.O. Box 400325, Charlottesville, VA 22904, USA}
\altaffiltext{12}{National Radio Astronomy Observatory, 520 Edgemont Road, Charlottesville, VA 22903, USA}

\begin{abstract}
We present an initial result from the $^{12}$CO~($J$=1--0) survey of 79 galaxies 
in 62 local luminous and ultra-luminous infrared galaxy (LIRG and ULIRG) systems 
obtained using the 45\,m telescope at the Nobeyama Radio Observatory.  
This is the systematic $^{12}$CO~($J$=1--0) survey of 
the Great Observatories All-sky LIRGs Survey (GOALS) sample.
The molecular gas mass of the sample ranges $2.2 \times 10^8 - 7.0 \times 10^{9}\,M_{\odot}$ 
within the central several kiloparsecs subtending $15\arcsec$ beam.
A method to estimate a size of a CO gas distribution is introduced, which is 
combined with the total CO flux in the literature.
The method is applied to a part of our sample and 
we find that the median CO radius is 1--4\,kpc. 
From the early stage to the late stage of mergers,
we find that the CO size decreases while 
the median value of the molecular gas mass in the central several kpc region is constant.
Our results statistically support a scenario where molecular gas inflows
towards the central region from the outer disk,
to replenish gas consumed by starburst, and that such a process is common in merging LIRGs.
\end{abstract}


\section{INTRODUCTION}
Luminous infrared galaxies \citep[LIRGs, infrared luminosity $L_{\rm IR(\lambda=8-1000\,\mu \rm m)} \geq 10^{11}\,L_{\odot}$,][]{SandersMirabel96}
are known to be very important for understanding the cosmic evolution of galaxies.
The contribution from LIRGs to the cosmic star formation rate density rises from $z\sim 0$ to 1, 
amounting to more than $\sim 50\,\%$ at $z \sim 1$ \citep{Elbaz02, LeFloch05, Caputi07, Magnelli09, Magnelli13, Goto10}.
Since LIRGs are powered by intense starbursts and can also harbor buried active galactic nuclei (AGN)
(e.g., \citealt{Armus07,Petric11}; \citealt[hereafter ST13]{Stierwalt13}; \citealt{Stierwalt14}),
they are excellent local sources in which to study the build up of stellar mass and the growth of central black holes 
in rapidly evolving galaxies.\defcitealias{Stierwalt13}{ST13}
The investigation of star formation in local LIRGs is an initial step towards understanding more distant LIRGs.

In local LIRGs, star formation is responsible for most of the infrared (IR) radiation \citep{Petric11}.
The contribution from AGN to the bolometric power output in LIRGs is low, 
but it is know to increase with IR luminosity (\citealt{Kim98, Petric11}; \citetalias{Stierwalt13}).
Local LIRGs are observed 
at a full range of galaxy-galaxy interaction and merger stages from pre-mergers to final stage mergers
(\citealt{Haan11, Petric11}; \citetalias{Stierwalt13}).
Therefore galaxy interactions/mergers, which can induce violent star-forming activities, 
are important events for the local LIRGs because they induce starbursts (SBs) and 
fuel the central black holes which may become AGNs.

Mergers and interactions can efficiently drive radial gas flows towards the galactic center \citep{BarnesHernquist96, MihosHernquist96},
and such fueling of gas into the central region is also important for igniting the SB or/and growth of the black-hole.
Many ultra-luminous infrared galaxies (ULIRGs, $L_{\rm IR} \geq 10^{12}\,L_{\odot}$) are often observed as advanced-stage mergers
(ULIRGs of early or mid-stage mergers are also found by \citealt{Dinh2001}).
In the advanced-mergers, concentrated molecular gas disks in the central sub-kpc region are found by interferometric observations 
\citep{DownesSolomon98, BryantScoville99, ImanishiNkanishi13, Xu14}.
This is consistent with the models where galaxy interactions and mergers drive radial flows of gas 
and accumulate gas in the central several kpc regions  \citep{BarnesHernquist96, MihosHernquist96, Iono04}.
In only two early stage mergers, the characteristic feature of inflowing molecular gas has been found \citep{Iono04, Iono05}.
It is generally difficult to observe direct characteristics of gas-inflows in interacting/merging galaxies,
because the timescale of inflow is as short as $10^8$\,yr \citep{Iono04} 
and the velocity fields are quite complicated \citep{Saito15}.
Therefore, it is still uncertain whether radial gas inflow is a ``common'' phenomenon through the merger sequence of LIRGs,
particularly even at intermediate stages of mergers.

Previous surveys exclusively employed small diameter telescopes with a large beam to derive the total flux of the whole galaxy.
\citet[hereafter GS99]{GaoSolomon99} presented CO observations of merging U/LIRGs with large beam telescopes 
(mainly NRAO 12m) 
and found that total CO luminosities decrease with decreasing projected separations between nuclei of merging galaxies.
This decrement is interpreted that the total molecular gas in merging U/LIRGs is depleted with advancing stages of mergers 
by their high star-forming activities.
In the meantime, if there are the common gas-inflows, 
in the central regions, the molecular gas mass of merging U/LIRGs are replenished by gas-inflows 
contrary to the gas-depletion by star formation.
Thus the molecular gas mass in the central regions should apparently increase or not change.
Studies on both the gas depletion in a whole galaxy and the gas supply in a central region help us to
understand the properties and evolution of merger-induced activities such as the distribution and contents of 
molecular gas at the late and post stages of mergers, and nuclear activities requiring the gas-fueling.
Selective measurements of gas in the central several kpcs in interacting/merging galaxies 
can provide information on gas that has undergone such inflow.
Such observations can be realized by large aperture single dish telescopes.
Of course, ALMA is already providing information on the sub-kpc gas content in local U/LIRGs \citep[e.g.,][]{Ueda14,Xu14,Xu15,Scoville14arxiv},
but single-dish telescopes can still provide valuable information on the gas properties of large samples of local LIRGs,
covering wide swaths of phase space and highlighting individual sources for ALMA follow-up 
(at least for those sources near or below the Celestial equator).
The beam size of the Nobeyama Radio Observatory 45\,m 
Telescope\footnote{Nobeyama Radio Observatory is a branch of the National Astronomical Observatory of Japan, National Institutes of Natural Sciences.} (NRO45), 
$15^{\prime\prime}$ at the frequency of CO(1--0), corresponds to a projected scale of $3.4$\,kpc radius at 94\,Mpc which is a median distance of our sample.
The central region of galaxies is also the site of intense activities characterizing local LIRGs, such as nuclear SB and/or AGN.
The mid-infrared (MIR) sizes of local LIRGs are indeed seen to be a few kpc in size, 
with this decreasing as the IR luminosity and dust temperature increase \citep{Diaz-Santos10,Diaz-Santos11}.
Selectively observing the central regions can help us understand the properties of the gas flow
and the fueling of the powerful central starbursts and AGN.

Local LIRGs are recognized to be rich in molecular gas from previous CO observations \citep{SandersMirabel96}.
Large surveys of IR bright galaxies in CO with single-dish telescopes, including local LIRGs 
and ULIRGs,
have been conducted extensively to date 
\citep{SandersMirabel85, Sanders86, Young86, Young89, Young95, Mirabel90, Sanders91, Mazzarella93, Downes93, Elfhag96, Solomon97, Curran00, Yao03, GaoSolomon04a, Saintonge11, Garcia12}.
In addition to their employed small-diameter telescopes which provide the total flux over the whole galaxy,
however, the sample selection criteria of these surveys are divided among observers,
e.g., biases towards IR- or CO-luminous sources \citep{GaoSolomon04a}.
This has prevented us from obtaining a homogeneous CO dataset of local LIRGs 
that spans a range in luminosity, energy source, and merger stage, 
which is required to infer on the general characteristics of LIRGs 
concerning cold molecular gas and star formation, in an unbiased manner.

Interferometric observations of CO emission lines have also been conducted 
for individual LIRGs.
\citet{DownesSolomon98} presented CO imaging of ten merging ULIRGs 
and found the concentrated molecular gas in their central regions.
\citet{BryantScoville99} performed CO imaging of five merging and two interacting LIRGs.
They also found massive and concentrated molecular gas in cores of mergers, but
found an extended distribution of molecular gas comparable to its optical disk in one interacting LIRG.
\citet{Gao97BIMA} observed ten LIRGs at various stages of mergers and
found a trend that, in their sample, the early merger-stage LIRGs show relatively large sizes of molecular gas
comparing to the advanced merging LIRGs.
Although the samples and interferometers are diverse between observations, 
these interferometric observations provide important knowledges to us regarding behavior of molecular gas in merging LIRGs.

In order to establish a large CO data set of local LIRGs using the NRO45 telescope,
we have conducted a CO survey of the northern sub-sample of the Great Observatories All-sky LIRG Survey \citep[GOALS,][]{Armus09} sample.
The GOALS sample is itself a flux-completed sub-set of the IRAS Revised Bright Galaxy Sample \citep[RBGS,][]{Sanders03} 
with $60\,\mu\rm m$ flux densities of more than 5.24\,Jy and Galactic latitude $|b|$ of more than 5 degree, 
and consists of 180 LIRGs and 22 ULIRGs.
In this paper of our CO survey, 
we present the results from CO observations of 79 individual galaxies in 62 GOALS LIRG systems. 
The CO was observed with a single beam towards the central regions corresponding to their brightest mid-infrared (MIR) position. 
Detailed results of mapping observations and studies of the relation between the cold molecular gas and the star formation rate (SFR) 
or the presence of AGN will be presented in a following paper (Paper II).

This paper is organized as follows. 
We describe the sample and the CO measurement in Section 2.
In Section 3, we present the CO spectra, the CO luminosity, the molecular gas mass, and the velocity width.
Our CO flux is compared with previous observations and an estimate of the extent of CO emitting regions is presented here.
In Section 4, we examine the mass and the extent of molecular gas along merger processes,
which give implications on merger-driven gas-inflow.
The summary and conclusion are given in Section 5.
Throughout this paper we adopt the cosmology parameters of $H_0 = 70\,\rm km\,s^{-1}$, $\Omega_{\rm m} = 0.28$, and $\Omega_{\Lambda} = 0.72$.

\defcitealias{Sanders91}{SSS91}\defcitealias{GaoSolomon99}{GS99}\defcitealias{GaoSolomon04a}{GS04}

\begin{figure}   
\epsscale{1}
\plotone{{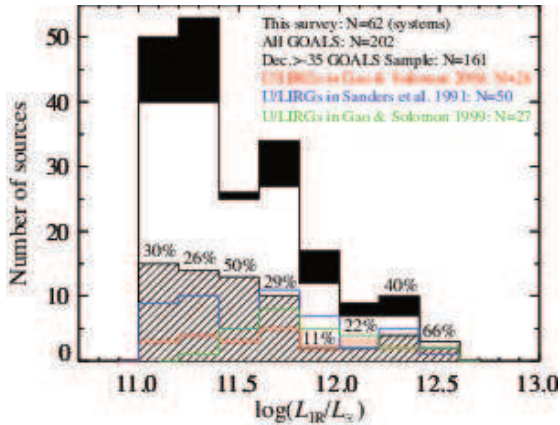}}
\caption{
The distribution of total IR luminosities $L_{\rm IR}$ in our NRO45 CO sample (shaded) and references: 
the full GOALS LIRGs (filled), the northern GOALS LIRGs (white), U/LIRGs in \citetalias{GaoSolomon04a} (red), 
U/LIRGs in \citetalias{Sanders91} (blue), and U/LIRGs in \citetalias{GaoSolomon99} (green).
The total IR luminosity is taken from \citet{Armus09}.
The numbers above each bin represent the fractional contribution of the CO sample to the full GOALS sample in that bin.
\label{fig:dist_lir}}
\end{figure}

\begin{figure}   
\epsscale{0.7}
\plotone{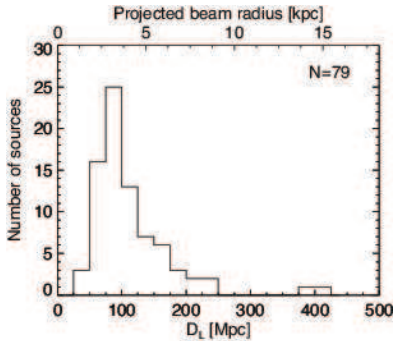}
\caption{The luminosity distances and the projected beam radii of $7.5^{\prime\prime}$ by NRO45 for the 79 individual sources.
The distances range from 37\,Mpc to 400\,Mpc corresponding to the beam radius of from 1.3\,kpc to 15\,kpc.
The median distance is 94\,Mpc, that is, the beam radius of 3.4\,kpc.
\label{fig:dist_distance}}
\end{figure}

\section{OBSERVATION}\label{sec:obs_and_reduction}

\subsection{Target Sample}\label{sec:sample}
Of the 202 full GOALS LIRGs \citep{Armus09}, 161 systems have declination of above $-35$ degrees, making them observable from the NRO45 (the Northern GOALS LIRGs).
Out of this sample, we have performed pointing-observations towards 79 galaxies in 62 systems (54 LIRGs and 8 ULIRGs), 
based solely on source availability at the assigned LST range and weather conditions.  
Here, we refer to a number of galaxies of mapping observations.
For 31 LIRGs whose MIR distributions ($8\,\mu\rm m$ of IRAC/{\it Spitzer}) are more extended than the beam of the NRO45, 
mapping observations have been performed.
Out of those, 11 galaxies are not overlap with the pointing-observation sample. 
Thus, a number of the total sample is 90 galaxies.

The IR luminosities of the 62 systems range from $L_{\rm IR}=1.10\times10^{11}\,L_{\odot}$ to $3.72\times10^{12}\,L_{\odot}$.
The number distribution of the systems of this sample with a bin of $\Delta\log{(L_{\rm IR}/L_{\odot})}=0.2$ is shown in Figure~\ref{fig:dist_lir}.
We compare this with three of the largest CO studies, 
\citet[hereafter SSS91]{Sanders91}, \citet[hereafter GS99]{GaoSolomon99}, and \citet[hereafter GS04]{GaoSolomon04a}.
\citetalias{Sanders91} conducted a CO survey for 50 local U/LIRGs with the NRAO 12\,m and the FCRAO 14\,m telescopes.
\citetalias{GaoSolomon99} presented CO luminosities of merging U/LIRGs with projected separations between nuclei.
Twenty-seven galaxies out of their entire sample belong to the IRAS RBGS.
\citetalias{GaoSolomon04a} conducted a survey of CO and HCN for 24 local U/LIRGs with IRAM 30\,m.
The four samples of \citetalias{Sanders91}, \citetalias{GaoSolomon99}, \citetalias{GaoSolomon04a}, and ours are subsets of the IRAS RBGS.
All the three comparing samples have the systematically higher IR luminosity range,
whereas our sample contains significantly more lower luminosity sources,
and a better representation of the entire GOALS sample.

Some LIRGs are known to be groups of two or more galaxies.
Our sample includes the 33 individual galaxies in 16 systems.
The individual IR luminosities of these galaxies are taken from \citet{Diaz-Santos10} and \citet{Howell10} 
who allocated the IR luminosities using the MIR flux density ratios.
For four pair galaxies (NGC~1797, IRASF10173+0828, NGC~4418, and MCG+04-48-002),
we estimate the individual IR luminosities in the same method using MIPS 24\,\micron\ flux density ratios.
Fifteen sources have the individual IR luminosities lower than $\log{(L_{\rm IR}/L_{\odot})}=11.0$
and drop out to ``sub-LIRGs'' \citep{Howell10}.
Using the conversion factor of \citet{Kennicutt98} from the IR luminosity to the SFR,
the individual SFRs are 0.07 -- 630\,$M_{\sun}\rm yr^{-1}$ and the median value is 36\,$M_{\sun}\rm yr^{-1}$ for all of the targeted sources.
For sources with $\log{(L_{\rm IR}/L_{\odot})} \ge 11.0$, the SFRs are 18 -- 630\,$M_{\sun}\rm yr^{-1}$ and the median value is 44\,$M_{\sun}\rm yr^{-1}$.
The heliocentric velocity and the luminosity distance of the sample are
taken from \citet{Armus09}.
The luminosity distances range from 37\,Mpc to 400\,Mpc ($z = 0.007-0.087$), and the median value is $94\,\rm Mpc$ (Figure~\ref{fig:dist_distance}).

\begin{figure}   
\epsscale{1}
\plotone{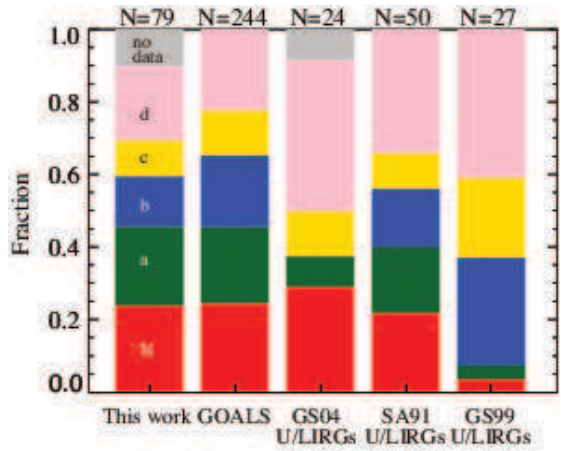}
\caption{
Fraction of merger stages for our sample (individual sources), the GOALS sample (individual sources), 
\citetalias{GaoSolomon04a} U/LIRGs, \citetalias{Sanders91} U/LIRGs, and \citetalias{GaoSolomon99} U/LIRGs.
Merger stages for the GOALS sources are taken from \citetalias{Stierwalt13}.
The merger stages are: N = non-merger (red), a = pre-merger (green), b = early stage merger (blue), c = mid-stage merger (yellow),
d = late stage merger (pink), as described in \citetalias{Stierwalt13}. 
``no data" means that a merger stage is unknown because of the accompanying sources not compile in \citetalias{Stierwalt13} or non-GOALS sample.
\label{fig:MS_histogram}}
\end{figure}

\begin{figure} 
\epsscale{1}
\plotone{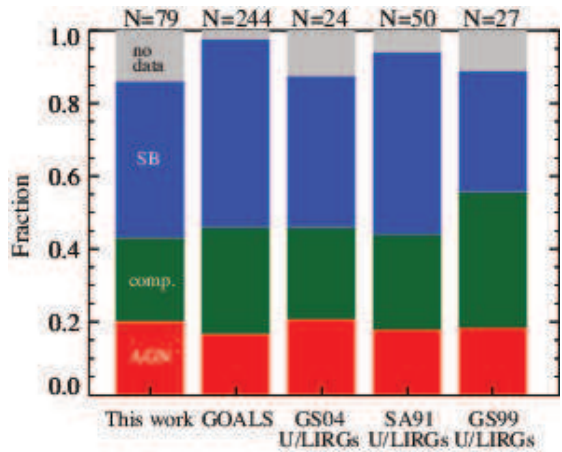}
\caption{Fraction of nuclear energy sources for our sample (individual sources), the GOALS sample (individual sources), 
U/LIRGs in \citetalias{GaoSolomon04a}, \citetalias{Sanders91}, and \citetalias{GaoSolomon99} U/LIRGs.
The sources of the GOALS sample are taken from \citetalias{Stierwalt13}.
The energy sources are distinguished into three groups according to the $6.2\mu\rm m$ PAH EQW in \citetalias{Stierwalt13}:
AGN (red), SB (blue), and composites of both (green). 
``no data" means that a PAH EQW is unknown in \citetalias{Stierwalt13}.
\label{fig:PAHEQW_histogram}}
\end{figure}

\citetalias{Stierwalt13} categorize the GOALS LIRGs into five stages of mergers: 
non-mergers with no sign of merger activities or massive neighbors, 
pre-mergers which are pair galaxies prior to its first encounter, 
early stage mergers which show symmetric disks but tidal tails, 
mid-stage mergers showing amorphous disks and tidal tails, 
and late-stage mergers with two nuclei in a common envelope (see \citetalias{Stierwalt13} for details). 
Our sample includes all the stages of the merger process, with a distribution similar to the GOALS sample (Figure~\ref{fig:MS_histogram}).
The U/LIRGs in \citetalias{GaoSolomon04a}, however, are biased towards late stage mergers.
In the sample of \citetalias{GaoSolomon99}, the number of galaxies at the non-merger and pre-merger is extremely low.
The nuclear energy sources are categorized into SB-dominated, AGN-dominated, and composite galaxies, 
in a manner identical to \citet{Petric11}, using the $6.2\,\mu\rm m$ polycyclic aromatic hydrocarbon 
equivalent widths (PAH EQW) in \citetalias{Stierwalt13}.
The dominant nuclear energy source of our sample is similar to the GOALS sample and U/LIRGs in \citetalias{Sanders91}, 
and also to \citetalias{GaoSolomon99} and \citetalias{GaoSolomon04a} which are limited in sample size (Figure~\ref{fig:PAHEQW_histogram}). 
Therefore, our sample is not biased towards specific phases of galactic interaction/merger 
or the dominant source of energy (SB or AGN).
The target sample is tabulated in Table~\ref{tab:source} with their properties and the observing coordinates.

\subsection{Nobeyama 45\,m Survey}
The $^{12}$CO ($J=1-0$) line emission was observed using the NRO45 over four semesters from January 2010 to February 2013.
The CO emission from the sources were measured with a single beam at the coordinates of the brightest points in their $24\,\mu\rm m$ of MIPS/{\it Spitzer} images.
The rest frequency of $^{12}$CO ($J=1-0$) emission is $115.27\,\rm GHz$.
The main beam size of the telescope at $115\,\rm GHz$ is $15\arcsec$.
The projected beam radius ranges 1.3 -- 15\,kpc (see Figure \ref{fig:dist_distance})
and the median value is $3.4\,\rm kpc$ at the median distance of the sample, $94\,\rm Mpc$.
Sixty-two objects ($78\,\%$) out of the sample have the projected radii of less than 5\,kpc.

The observations employed two types of dual-polarization two sideband-separating SIS mixer systems, 
T100H/V \citep{Nakajima08} and TZ \citep{Nakajima13}, as a front-end, depending on availability.
As the backend spectrometer in the first run, the wide-band-mode acousto-optical spectrometer (AOS-W) and the digital autocorrelator (AC45) were employed.
AOS-W and AC45 have the frequency bandwidths of $250\,\rm MHz$ and  $512\,\rm MHz$, corresponding to $\sim 652\,\rm km\,s^{-1}$ and $1336\,\rm km\,s^{-1}$ velocity bandwidths at $115\,\rm GHz$, respectively. 
The frequency resolutions are $250\,\rm kHz$ and  $910\,\rm kHz$, corresponding to velocity resolutions $0.652\,\rm km\,s^{-1}$ and $2.37\,\rm km\,s^{-1}$, for AOS-W and AC45 respectively.
From the second run to the fourth, we utilized a broad bandwidth spectrometer, SAM45 \citep{Kamazaki12}, which enables us to obtain spectra in a broader bandwidth than those in the previous run.
We used SAM45 with a frequency coverage of $2\,\rm GHz$ (=$5217\,\rm km\,s^{-1}$) and a frequency resolution of $488\,\rm kHz$ ($=1.27\,\rm km\,s^{-1}$).
For some galaxies whose spectra were obtained with several spectrometers, the one with the higher signal-to-noise (S/N) ratio is used for analysis.
Consequently, eighteen galaxies are observed by AOS-W, six by AC45, and fifty-five by SAM45 (see also Table \ref{tab:COresult}).

Fluxes of the obtained spectra were calibrated with the chopper wheel method and by observing standard sources.
System noise temperatures, throughout all the observations, were typically $150-250\,\rm K$ on the antenna temperature scale ($T^*_{\rm A}$) at the observed frequency.
The pointing of the telescope was regularly checked every 30 - 60 minutes by observing SiO masers, with the typical r.m.s. error of $\sim 3''$.
The spectra of observed CO emission lines are shown in Figure \ref{fig:spectra} with the observed position of the NRO45 beam, 
superposed on the 3.6\,$\mu$m images and 24\,$\mu$m contours.

\subsection{Data Reduction and Analysis}
The primary data reduction was performed by {\it NewStar}, a software distributed by NRO.
The baselines were fitted with polynomial functions of order 1,
but we used second to third orders in some cases that show large baseline fluctuations.
Averaging individual scans of a source was done using weights of $1/\mathrm{rms}^2$.
The spectra were then smoothed with a box function of a velocity resolution of $20\,\rm km\,s^{-1}$ in order to improve the S/N ratio, except for IRAS~F12224-0624 smoothed to a bin of 60\,km\,s$^{-1}$ and Mrk~331 to a bin of 40\, km\,s$^{-1}$.
The CO velocity-integrated intensity was derived from the main beam temperature
and the full velocity width and is represented as follows:
\[
I_{\mathrm{CO}} = \int T_{\mathrm{mb}}\, dV  =  \int \frac{T^*_{\mathrm{A}}}{\eta_{\mathrm{mb}}}\, dV\, ,  
\]
where $I_{\mathrm{CO}} $ is in a unit of $\rm K\,km\,s^{-1}$, 
$T_{\rm mb}$ is the main beam temperature in K and 
$\eta_{\rm mb}$ is the main beam efficiency which is 0.38, 0.36, 0.31, and 0.28\footnote{The main beam efficiency is taken from the NRO web page http://www.nro.nao.ac.jp/~nro45mrt/html/ .}
in the first, second, third, and fourth runs, respectively.
The performed integral range is indicated by a horizontal solid line below the spectrum in Figure \ref{fig:spectra}.
In order to estimate systematic calibration uncertainties, we further calibrated the integrated intensity using the observations of standard sources, NGC~7538, IRC+10210, and W51.
The calibration factor is determined for each observing day and ranges from 0.838 to 2.80.
The median is 1.14.
Standard sources were not observed in the first run, so were bootstrapped to sources which were observed over several runs.
The uncertainties in $I_{\rm CO}$ which is estimated from variations of fluxes of standard sources are typically about 13\,\%.

Of the 79 observed  sources, 68 sources were detected, giving a detection rate of 86\%.
Five out of 11 non-detected sources are companions with fainter IR luminosity in their paired system.
Three objects out of eleven non-detections were observed by AOS-W. Therefore, the narrow bandwidth is not 
a major cause of the non-detections.
For the non-detected sources, 
3 $\sigma$ upper limits of their integrated CO intensity were estimated on the assumption that the sources have 
a Gaussian-shaped spectrum with an average velocity width of the sample, $300\,\rm km\,s^{-1}$.
Additional information for the notable sources are presented in Appendix \ref{apx:note_obj}.

\section{RESULTS}\label{sec:result}
\subsection{CO Luminosity and Molecular Gas Mass}\label{sec:COandMH2}
\begin{figure}
\epsscale{0.9}
\plotone{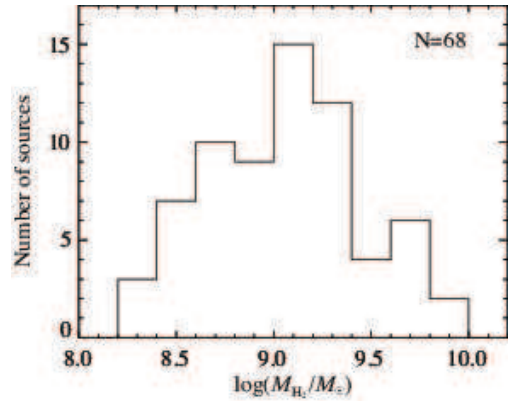}
\caption{The histogram of the molecular gas mass of the sample. 
We adopt $alpha_{\rm CO}$ of $0.6\,M_{\odot}(\rm Kkm\,s^{-1}pc^{-2})^{-1}$ \citep{Papadopoulos12} to calculate molecular gas mass.
The average is  $1.70 \times 10^{9}\,M_{\odot}$.
\label{fig:hist_mass}}
\end{figure}

The CO luminosity is commonly described by the following equation: 
\[
L^{\prime}_{\mathrm{CO}} = I_{\mathrm{CO}} \Omega_{\mathrm{beam}} D_{\mathrm{L}}^2 (1+z)^{-3}\, ,
\]
where $\Omega_{\rm beam}$ is the main beam solid angle,
and $D_{\rm L}$ is the luminosity distance.
The molecular gas mass, $M_{\rm H_2}$, was calculated using a CO-H$_2$ conversion factor, 
$\alpha_{\rm CO}  = M_{\rm H_2}/L^{\prime}_{\rm CO}$.
In this paper, an $\alpha_{\rm CO}$ of $0.6 \pm 0.2 \,M_{\odot}(\rm Kkm\,s^{-1}pc^{-2})^{-1}$ \citep{Papadopoulos12} is used for all sources in the sample.
This value is taken from \citet{Papadopoulos12} who performed a one-phase radiative transfer analysis using global $^{12}$CO and $^{13}$CO emission lines of local ULIRGs and LIRGs. 
Such low values of $\alpha_{\rm CO}$ compared to the Galactic value of  $\sim 4.3 \,M_{\odot}(\rm Kkm\,s^{-1}pc^{-2})^{-1}$ 
are generally accepted for ULIRG or LIRGs (e.g., \citealt{DownesSolomon98}; \citealt[and references therein]{Bolatto13}).
The $\alpha_{\rm CO}$ is known to be affected by some parameters, including metallicity, surface gas density, and dust temperature \citep[e.g.,][]{Tacconi08, Genzel12, Magnelli12}.
However, within the population of local LIRGs, the dynamic range of these parameters are relatively small,
and thus these parameters are likely to be ineffective in changing the $\alpha_{\rm CO}$ in our LIRG sample.
For instance, the dust temperatures of our sample, which are derived from the IRAS 60\,$\mu$m/100\,$\mu$m color and $\beta$ of 2,
are in a narrow range of 30 -- 40\,K.
This corresponds to a variation of a factor of three on the relation between the dust temperature and $\alpha_{\rm CO}$ 
derived by \citet{Magnelli12}.
In addition, the dispersion of the observed CO intensities is as small as $\sim$0.5\,dex in our sample, which provides a small variation of less than a factor of two based on the correlation between the total surface density and $\alpha_{\rm CO}$ \citep{Bolatto13} assuming that local LIRGs have the same gas fraction.
\citet{Inami13} has shown that the gas-phase metallicities in GOALS sources are
confined to a relatively narrow range of $1 < Z(Z_{\odot}) < 2$.
This metallicity range corresponds to the nearly constant $\alpha_{\rm CO}$ 
according to the predictions in \citet{Bolatto13} using the models by \citet{GloverMacLow11} and \citet{Wolfire10}.
We do not consider the variation of $\alpha_{\rm CO}$ within a galaxy \citep{Sandstrom13}, 
because only the central regions are observed.
We note that the comparisons of molecular gas mass between LIRGs are less affected by
the systematic variation of $\alpha_{\rm CO}$ over the populations of galaxies as long as we use a single $\alpha_{\rm CO}$.
The calculated CO luminosity and molecular gas mass are shown in Table \ref{tab:COresult}.
The molecular gas masses range from $2.2\times 10^8\,M_{\odot}$ to $7.0\times 10^{9}\,M_{\odot}$, and 
the histogram of $M_{\rm H_2}$ is shown in Figure \ref{fig:hist_mass}.

\subsection{Velocity Width}
\begin{figure}
\epsscale{0.9}
\plotone{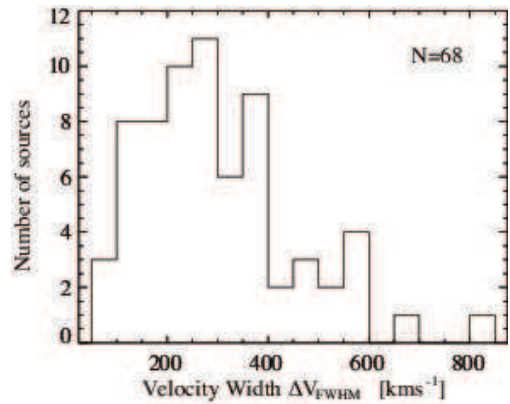} 
\caption{The histogram of the velocity width $\Delta V_{\rm FWHM}$ of CO line. 
The average value is 300\,km\,s$^{-1}$. The maximum is 840\,km\,s$^{-1}$ for UGC 05101.
\label{fig:hist_dV}}
\end{figure}

We derived the full width at half maximum (FWHM) of the velocity width, $\Delta V_{\rm FWHM}$, of the CO line of all detected sources.
The error in the $\Delta V_{\rm FWHM}$ is estimated by its variation which depends on the peak intensity uncertainty, and is $\sim 35$\,\% on average.
The distribution of $\Delta V_{\rm FWHM}$ is shown in Figure \ref{fig:hist_dV}.
We find that $\Delta V_{\rm FWHM}$ is $\sim 300$\,km\,s$^{-1}$ on average, ranging from 80\,km\,s$^{-1}$ to 840\,km\,s$^{-1}$.
Two galaxies, UGC~05101 and MCG~+04-48-002S, have the significantly high velocity-widths 
(839\,km\,s$^{-1}$ and 651\,km\,s$^{-1}$, respectively)
over observed ones by \citetalias{GaoSolomon04a}, 245\,km\,s$^{-1}$ and 397\,km\,s$^{-1}$, respectively.
This may be because our observation using the broader spectrometer
could capture components of gas with anomalous velocities.
While the maximum velocity coverages of the employed spectrometer is $\sim 5200\,\rm km\,s^{-1}$ at 115\,GHz, 
those in \citetalias{Sanders91} and \citetalias{GaoSolomon04a} are 1330\,km\,s$^{-1}$ and 1560\,km\,s$^{-1}$, respectively.

\subsection{Comparison with Results of the Previous Observations}\label{sec:comparison_flux}
We compare the CO fluxes of 24 individual sources that are in both our observation and previous observations
(\citetalias{GaoSolomon04a}, \citealt{Solomon97}, \citealt{Young95}, and \citetalias{Sanders91}),
which used smaller diameter telescopes of 12\,m, 14\,m, and 30\,m 
with beam sizes of 55$^{\prime\prime}$, 45$^{\prime\prime}$, and 22$^{\prime\prime}$, respectively.
In case of multiple entries in the literature, we use values from smaller aperture telescopes, 
and in case more than one value is found for a telescope with the same aperture, 
we list the result from the most recent observation.
For UGC~08387, the value from FCRAO 14\,m \citep{Young95} is used as a reference instead of NRAO 12\,m \citepalias{Sanders91} 
for the purpose of the CO size estimation in Section \ref{sec:co_estimate}, 
as the 12\,m measured a flux that was smaller than the 45\,m measurement.

\begin{figure}
\epsscale{1}
\plotone{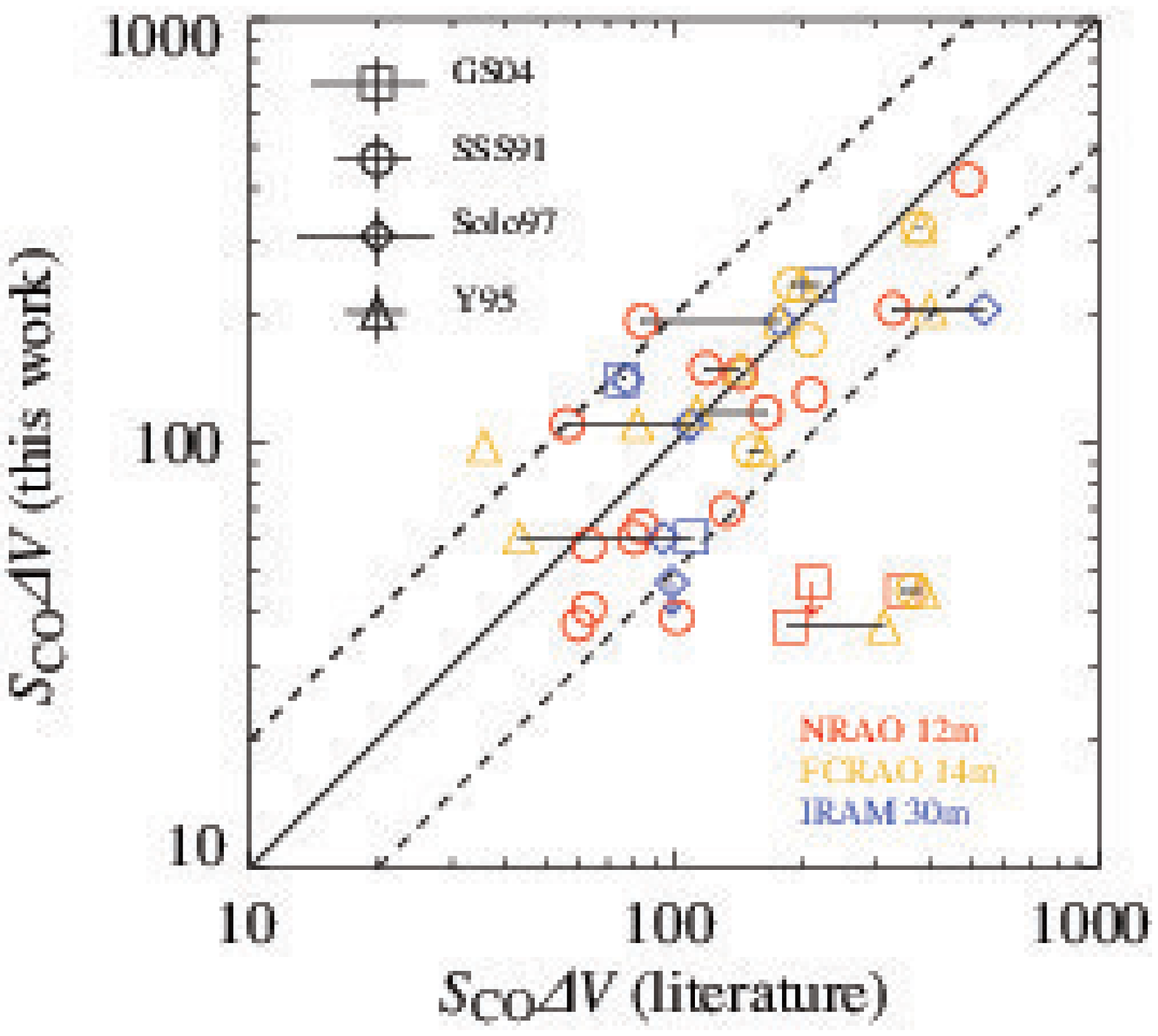}
\caption{
The comparison of the CO fluxes of 24 sources that include in both our NRO45 observations with the literature
which are \citetalias{GaoSolomon04a} (square), \citetalias{Sanders91} (circle),
\citet{Solomon97} (diamond), or \citet{Young95} (triangle).
The color represents the employed telescope: NRAO 12\,m (red), FCRAO 14\,m (orange), and IRAM 30\,m (blue).
The typical errors in each observation are shown on the upper-left.
The arrow means an upper limit of flux.
The horizontal solid line is connecting each point of an identical galaxy but different observations.
The solid and dotted lines represent where the flux of this work is equal to the flux of the literature and is shifted by a factor of 2, respectively.
\label{fig:IcoA_comp}}
\end{figure}

\begin{figure}
\epsscale{1}
\plotone{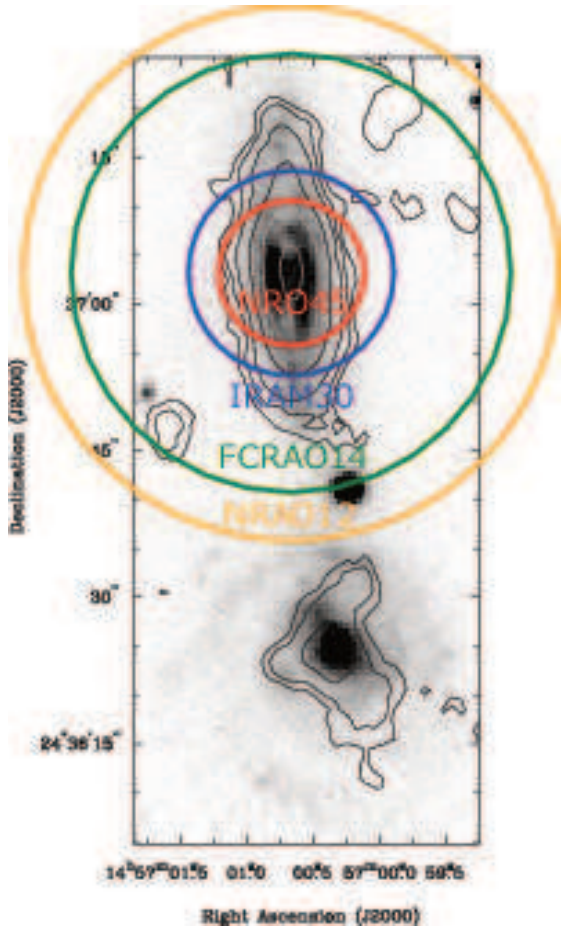}
\caption{
The illustration of the effect of the different beam sizes on the total CO flux.
The contour shows integrated CO(1--0) intensity map of a LIRG Arp~302, which is taken from \citep{Lo97}.
The gray scale is an R-band image. 
The four circles indicate the CO beam sizes of NRO 45\,m (red), IRAM 30\,m (blue), FCRAO 14\,m (green) and NRAO 12\,m (red).
\label{fig:beams}}
\end{figure}

The comparison of the CO fluxes between this work and the references is shown in Figure \ref{fig:IcoA_comp}. 
The fluxes of this work are on average $16\,\%\pm14\,\%$ lower than those of the NRAO 12\,m observations while
they are on average nearly equal to those of the IRAM 30\,m observations.
The fluxes of our observation are on average $23\pm14\,\%$ lower 
than those of the FCRAO 14\,m observations if we exclude 3 ULIRGs and NGC 1275, 
whose $\Delta V_{\rm FWHM}$ is substantially different between our observation and \citet{Young95} (see Appendix \ref{apx:note_obj} for details).
This systematic decrease is because of the difference of the projected beam sizes of the telescopes.
The projected radii of the NRO45 and the 30\,m telescope are 3.5\,kpc and 5.2\,kpc at 97\,Mpc, respectively.
The beams can capture the flux from only the central region.
On the other hand, the projected radii of the 12\,m and 14\,m telescopes are
13\,kpc and 11\,kpc at 97\,Mpc, respectively.
Because our Galaxy has an extended CO distribution up to a radius of $\sim 10\,\rm kpc$ \citep{NakanishiSofue2006} and
the spiral galaxies in the Virgo cluster have the CO emitting regions of radii up to 13.3\,kpc \citep{Nakanishi2005},
the beams of the 12\,m and 14\,m telescopes are large enough to observe the total flux of a galaxy.
The fluxes between NRAO 12\,m and FCRAO 14\,m are not significantly different.
We illustrate the effect of the different beam sizes in Figure \ref{fig:beams}.
On a CO contour of LIRG Arp~302 taken from \citet{Lo97}, four beam sizes are shown.
This example obviously shows NRO45 and IRAM 30\,m beam sizes are smaller than the extent of CO gas
while NRAO 12\,m and FCRAO 14\,m beams are enough large.
It is worth mentioning that the extended molecular gas of some LIRGs have been reported from 
CO(3--2) mappings with the JCMT telescope whose beam size is equivalent 
to one of NRO45 \citep{Leech10}. 
This strongly implies that some LIRGs have the extended CO(1--0) distributions beyond the NRO45 beam.

\subsection{CO Size Estimation}\label{sec:co_estimate}
Molecular gas distributions in local LIRGs can be measured by interferometric observations, 
which show that almost all molecular gas of advanced mergers is concentrated in the compact nuclear region \citep{DownesSolomon98, BryantScoville99}
while molecular gas of some early or mid-stage mergers extends over a disk or a overlap region \citep{Lo97,Gao97,Wang01,Saito15}.
However, it is time-consuming for interferometers to build up a large sample of CO distributions of merging LIRGs.
A method of comparisons of our NRO45 flux with observations from the literature which used single-dish telescopes with larger beam sizes
allows us to quickly estimate the spatial extent of the CO distribution of our sources 
and to produce a large unbiased sample of galaxies with an estimated extent of CO gas.

The size of the CO distribution, $\mu$, can be estimated by using the flux ratio of
our flux at the NRO45 to the flux at the 12\,m or 14\,m telescopes.
The flux ratio is an observable value, and represents the ratio of the the central flux 
convolved with the NRO45 beam to the total flux, 
on the supposition that the flux from the 12\,m or 14\,m observations are the total flux of a galaxy
due to their large beam sizes (see Section \ref{sec:comparison_flux}).
The flux ratio is rewritten as,
\begin{equation}\label{eq:Rco}
R_{\rm CO}=\frac{S_{\rm CO}dV(\rm{NRO45})}{S_{\rm CO}dV(\rm{ref})}= \frac{\int I(\mbox{\boldmath $r$}, \mu) P_n(\mbox{\boldmath $r$})\,d\mbox{\boldmath $r$}}{\int I(\mbox{\boldmath $r$}, \mu)\,d\mbox{\boldmath $r$}}
\end{equation}
where $I(\mbox{\boldmath $r$},\mu)$ is the radial CO intensity distribution which is characterized by the size $\mu$,
$\mbox{\boldmath $r$}$ is the radius from the center of the galaxy on the sky,
$P_n(\mbox{\boldmath $r$})$ is the normalized beam pattern of NRO45, 
and the integral is performed over the whole galaxy. 
The radial distribution of CO intensity can often be approximated by an exponential, 
a Gaussian, or an uniform-disk distribution \citep[e.g.,][]{Young95,Nishiyama01b}. 
We assume these three profiles of the CO intensity, 
the Gaussian distribution (Model~A), the azimuthally symmetric exponential distribution (Model~B), 
and the uniform disk (Model~C).
The size of the CO extent $\mu$ is defined as the half width at the half maximum (HWHM) 
of the CO distribution for the model~A ($\mu_A$) and Model~B ($\mu_B$),
and the radius of the CO distribution for the Model~C ($\mu_C$).
In brief, this method solves Equation \ref{eq:Rco} for the parameterized CO size $\mu$
using the observed flux ratio $R_{\rm CO}$ and the assumption of the CO intensity distribution $I(\mbox{\boldmath $r$},s)$.
Note that there is a limit to this estimate for galaxies with off-center distributions of CO gas.
Moreover, there may be possibilities of overestimations for some galaxies 
whose CO distributions are extended up to 12\,m or 14\,m beams.
For example, the Antennae galaxies show the extended molecular gas comparable to its optical distribution \citep{Gao01}.
The optical sizes of our sample galaxies are as well as or slightly larger than 12\,m and 14\,m beams.
Further description of the estimation is given in Appendix \ref{apx:COsize_model}.

Sizes of the CO distribution of a sub-sample are estimated using this method.
The sub-sample is composed of 21 galaxies 
which have literature flux value of the 12\,m or the 14\,m telescopes 
and have the value of $R_{\rm CO}$ less than one within the $1\sigma$ uncertainty.
For four galaxies with $R_{\rm CO} \geq 1$ we cannot estimate $\mu$ from $R_{\rm CO}$ directly,
and therefore we estimate only the $1 \sigma$ upper limit of $\mu$ 
by replacing $R_{\rm CO}$ in Equation \ref{eq:Rco} by $R_{\rm CO} - 1 \sigma$.
The flux ratio $R_{\rm CO}$ and the CO size $\mu$ of the sub-sample are summarized in Table \ref{tab:COsize}.

\begin{figure}
\epsscale{1}
\plotone{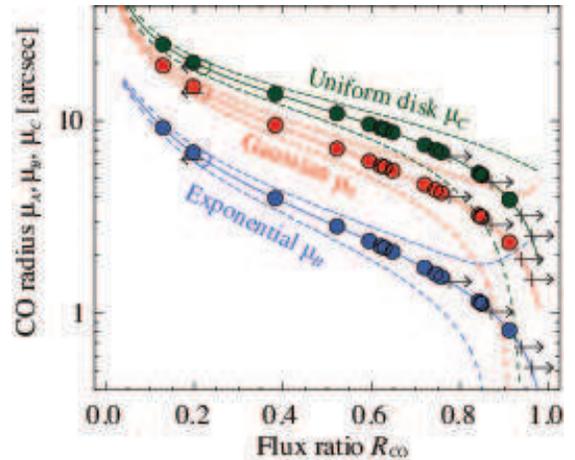}
\caption{
The modeled angular CO radius, $\mu$, 
and the CO flux ratio, $R_{\rm CO}$, 
which is the ratio of our NRO45 observation to the large beam observation with the 12\,m or 14\,m telescopes,
of the sub-sample which can be estimated.
The CO radius $\mu$ in y-axis means $\mu_A$ for the Gaussian model (Model~A, red),
$\mu_B$ for the exponential model (Model~B, blue), and $\mu_C$ for the radius for the uniform disk model (Model~C, green).
The solid and the dashed lines indicate the model curve of A, B, and C, respectively.
The right (left) arrow denotes the lower limit (upper limit) of $R_{\rm CO}$ and the upper limit (lower limit) of $\mu$.
\label{fig:R_COsize}}
\end{figure}

Figure \ref{fig:R_COsize} shows $\mu$ as a function of $R_{\rm CO}$ for each of the models.
The difference of $\mu$ among three models increases from a factor of 2 around $R_{\rm CO}$ of 0.1, 
up to a factor of 5 near $R_{\rm CO}$ of 1.
The error of each model also increases with $R_{\rm CO}$,
becoming greater than 100\,\% at $R_{\rm CO} \gtrsim 0.85$. 
The source IRASF~01417+1651 in all models, and IC~1623AB, NGC~0992, and NGC~7771S1 in Model B 
have the errors of more than 100\,\% in the estimated CO size.
In other words, the accuracy of this method decreases rapidly
for $\mu$ of $\lesssim 2\arcsec$, $\lesssim 1\arcsec$, and $\lesssim 4\arcsec$ in Model~A, B, and C, respectively.
Therefore, this method is not sensitive to CO distributions that are much smaller than the beam size of the single-dish observations.

\begin{figure}
\epsscale{1}
\plotone{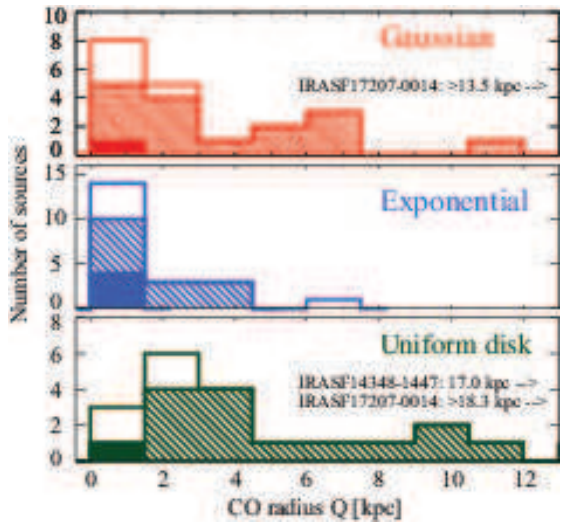}
\caption{
The CO radius $Q$ in unit of kpc of the sub-sample.
The three models are shown, the Gaussian model (red), 
the exponential model (blue), and the uniform disk model (green).
The values of the upper limit and the lower limit are also shown by the bland box and the shaded box, respectively.
The sources with a large error over 100\,\%, which are marked with the colon in Table \ref{tab:COsize}, are showed as the filled box. 
IRAS F14348-1447 of the Gaussian and uniform models, and IRAS F17207-0014 of the uniform model
are out of the range in the lower diagram
due to their large radius of 13.5\,kpc, 17.0\,kpc and $>18.3$\,kpc, respectively.
\label{fig:actCOsize}}
\end{figure}

The ranges of CO size of the sub-sample are 
2\farcs3 -- 19\farcs4, 3\farcs7 -- 9\farcs8, and 3\farcs9 -- 25\farcs0 in Model~A ($\mu_A$), B ($\mu_B$), and C ($\mu_C$), respectively.
Eight sources (38\,\% of the sub-sample) show compact distributions in Model~B with $\mu_B \lesssim 1$.
Figure \ref{fig:actCOsize} shows the CO radius on the physical scale $Q$ of the three models.
The $Q$ widely ranges from 0.8\,kpc to 11\,kpc in Model~A and from 1.3\,kpc to 17\,kpc in Model~C.
Meanwhile all $Q_B$ in Model~B settle within a narrow range from 0.3\,kpc to 4\,kpc.
The median $Q$ of Model~A, B, and C are 2.6\,kpc, 1.0\,kpc, and 4.1\,kpc, respectively, 
excluding upper/lower limits.
The majority of the sources, 14 sources (67\,\%) for Model~A, 20 sources (95\,\%) for Model~B, and 13 sources (62\,\%) for Model~C, 
have $Q$ of less than 4\,kpc.

We compare $\mu$ with sizes measured with interferometers, to verify this method.
We summarize published interferometric data of CO~(1--0) and CO~(2--1) in Table \ref{tab:COsize}. 
In nine of twelve galaxies (75\,\%) that have the interferometric measurements in literature, 
the estimated sizes in one or more models
are consistent with interferometric measurements within the errors.
The $\mu_A$ is consistent with interferometric results for seven galaxies (58\,\%).
This ensures the validity of this size-estimation.
The sizes of two sources, NGC~4418, and IRAS~F17207-0014, are inconsistent with the interferometric measurements.
These sources have a very compact distribution of $<1^{\prime\prime}$, for which our method is likely invalid.

\begin{figure}
\epsscale{1}
\plotone{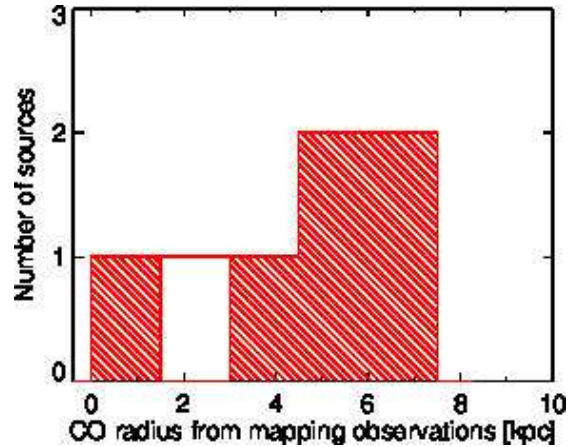}
\caption{
The CO size measured with the mapping observations.
The size is define as the beam-deconvolved HWHM of a fitted Gaussian to a radial profile of an integrated intensity (see Appendix \ref{apx:mapping}).
The shaded box is the measured CO size, and the blank box is the upper limit of the size.
\label{fig:map_cosize}}
\end{figure}

In addition to the comparison with the interferometric sizes, the estimated CO sizes are compared with CO sizes measured 
from our mapping observations.
Although no galaxies overlapping between the two measurements, 
this comparison allows us to confirm that our estimate does not largely contradict.
The CO sizes from the mapping observations are measured for seven galaxies out of the mapping sample.
These galaxies have CO detections in more than three positions along a axis and 
the peak positions of the intensities do not deviate from the galactic center.
The CO size from the mapping observations is defined as a beam-deconvolved HWHM of a Gaussian fitting to a radial profile of an integrated intensity,
and is summarized in Table \ref{tab:mapping}.
The detailed method to measure the size and an example of the mapping observations are presented in Appedix \ref{apx:mapping}.
Figure \ref{fig:map_cosize} shows that the CO size in kpc scale from the mapping observations distributes from 1 to 7\,kpc.
This range is comparable to one of the Gaussian model in Figure \ref{fig:actCOsize}.
Therefore we conclude that our estimate of the CO size based on two different telescopes is consistent with the measurements of the mapping observations.

For the Gaussian distribution, $Q_A$ corresponds to a radius where a half of total energy is included.
Our Galaxy has 50\,\% of its molecular gas mass within the radius of $\sim$6\,kpc \citep{Sanders84}.
The CO distributions of almost all of our sources are concentrated in a region more compact compared to our Galaxy.
\citet{YoungScoville1982} reported that the HWHM of two late-type spiral galaxies 
whose radial profiles are fitted well by an exponential function 
which is similar to Model~B is estimated to be approximately 4\,kpc.
\citet{Nishiyama01b} fitted CO radial distributions in the outer disk regions of nearby spiral galaxies by an exponential function
and estimate a HWHM of $2.3\pm1.5$\,kpc on average.
The median radius of $Q_B$ (1.0\,kpc) is approximately a factor of 2-4 lower than spiral galaxies.
The $Q_B$ of 17/21 sources are smaller than the spiral galaxies.

For sources whose interferometric data are unavailable, 
the size estimate based on the Gaussian model predicts 
NGC~0992 and NGC~7771S1 to possess a compact distribution of the cold molecular gas ($\lesssim 1\,\rm kpc$).
This could imply the presence of the nuclear molecular disk in these sources.
Interferometric observations of some local LIRGs reveals that
the molecular gas is concentrated towards the central 1\,kpc \citep[e.g.,][]{DownesSolomon98,BryantScoville99}.
On the other hand, our model also predicts that IRAS~F03359+1523, IC~2810W, MCG+04-48-002S, and Mrk~331 
have molecular gas distribution extended over $\sim 5$\,kpc.
This method can easily estimate rough sizes of CO distributions.

\section{DISCUSSION}

\subsection{Molecular Gas Inflow in Merging Galaxies}\label{sec:merger_gas}
\subsubsection{Molecular Gas in the Central Region and Merger Process}
It is difficult to observe inflowing gas in interacting galaxies
because of the short time-scale of inflow \citep[$10^8\,\rm yr$,][]{Iono04}
and complicated morphologies and velocity fields in mergers.
In this section, we assess whether gas inflow is a common process in local LIRGs.
\citet{GaoSolomon99} find that the {\it total} molecular gas mass of LIRGs and ULIRGs 
decreases by a factor of $\sim$ 5 with decreasing separation between the two nuclei.
This decrease in mass is interpreted as the consumption of molecular gas by the interaction/merger induced starburst.
This is consistent with theoretical predictions in which 
$\sim 70\,\%$ of the initial gas content is consumed in a major merger \citep{Cox08}.

Our observations allow us to statistically investigate the supply and consumption of the gas 
in the central region ($r \sim 3.4\,\rm kpc$, see Figure \ref{fig:dist_distance}) along the merger sequence.
While the total molecular gas in the whole galaxy decreases due to star formation, 
if molecular gas inflow from the outer parts of the galaxy into the central region is common in our sample, 
we expect the molecular gas in the central region to either be increasing when a speed of gas inflow is higher than
one of gas-consumption by star formation, or be constant when the both speeds are equivalent,
sustaining the high SFR during the interaction/merger.

We classify our sources into three stages of merger instead of the five stages in Section \ref{sec:sample}
in order to obtain the statistically large number of sources at each stage.
The original five stages are binned into the three stages as follows:
the non-merger (N) remains as the non-merger (0), 
the original pre-merger (a) and early stage merger (b) are combined as early stage (1), 
and original mid-stage merger (c) and late stage merger (d) are also combined as late stage (2).
The numbers of galaxies with $\log{(L_{\rm IR}/L_{\odot})} \geq 11.0$ in each merger stage are
15, 21, and 22 for the non-merger, the early stage merger, and the late stage merger, respectively.
\citet{Haan11} investigated the merging timescales of a sub-sample in the GOALS sample 
and reported a timescale of 0.3 -- 1.3\,Gyr until the nuclei merge for interacting GOALS galaxies.
Thus we can estimate the dynamical timescale between the early stage and late stage to be approximately $1\,\rm Gyr$.

\begin{figure}
\epsscale{1}
\plotone{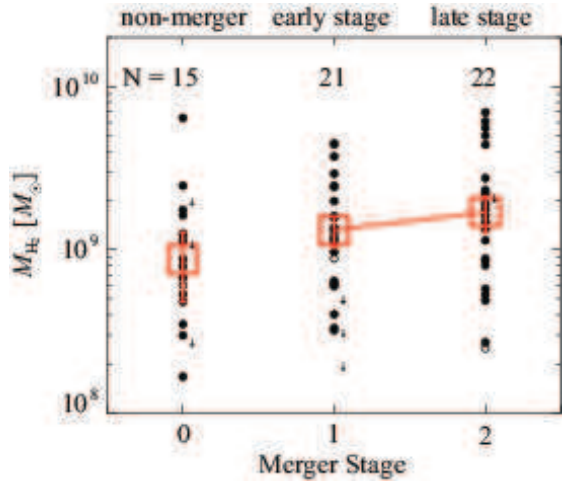}
\caption{
The molecular gas mass in the central region at each merger stage.
The sources are grouped into 3 stages of merger: `0' = non-merger, `1' = early stage, `2' = late stage.
The filled and open circles mean the sources with $\log{(L_{\rm IR}/L_{\odot})}\ge 11.0$ and $\log{(L_{\rm IR}/L_{\odot})}< 11.0$, respectively.
The arrows represent the non-detection sources with the upper limits of the molecular gas mass.
The red squares indicate the median values of the detected sources with $\log{(L_{\rm IR}/L_{\odot})}\ge 11.0$ at the each stage, whose numbers of sources are shown on the upside of the diagram.
\label{fig:merger_mass}}
\end{figure}

\begin{figure}
\epsscale{0.8}
\plotone{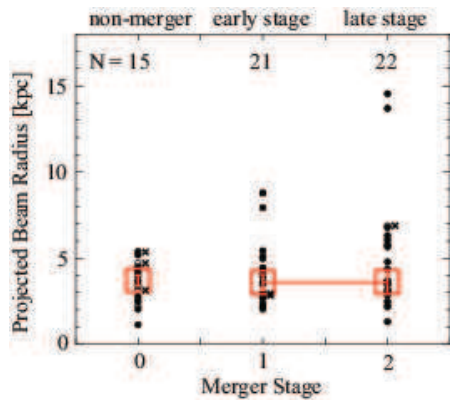}
\caption{
The projected beam radius of the sample at each merger stage.
The symbols are the same as Figure \ref{fig:merger_mass} but the crosses indicate the non-detection sources.
\label{fig:merger_beam}}
\end{figure}

Figure \ref{fig:merger_mass} shows the molecular gas mass in the NRO45 beam at each merger stage.
In this diagram, we also plot the median values among sources detected
in CO with $\log{(L_{\rm IR}/L_{\odot})} \geq 11.0$.
We do not detect any trend in the molecular gas mass in the central region along the merger sequence.
Although there is a huge dispersion and therefore any quantitative discussion is difficult,
the median values of the molecular gas mass from the early stage to the late stage are constant within their errors.
The result of a Kolmogorov-Smirnov (K-S) test indicates a $p$-value of 0.66 
between the distributions of the early stage and the late stage,
which suggests that the molecular gas mass in the central region, on average, 
does not evolve strongly as a function of merger stage,
and supports the constant molecular gas mass in the central region.
The {\it total} molecular gas mass decreases by a factor of $\sim$ 5 
along the merger sequence as proposed by \citet{GaoSolomon99}, 
while but the molecular gas mass in the central region appears not to show any similar trends.
This result could imply a radial inflow towards the central region from the outer disk.

The molecular gas mass in the central region of the NRO45 beam along the merger stage 
may be affected by a systematic bias in the the source distance as the function of the merger stage,
since the area subtended by the telescope beam would be changed.
Figure \ref{fig:merger_beam} shows the distribution of the projected radii at each stage.
Almost all of the sources are within 2\,kpc -- 5\,kpc,
and no systematic dependence of the projected radii on the merger stage is present.
Even if four sources with a particularly large projected radius of $\geq 7\arcsec$ 
are excluded from Figure \ref{fig:merger_mass}, 
the median of the molecular gas mass still does not detect any trend.

A systematic dependence of $\alpha_{\rm CO}$ on the merger stage can also affect the molecular gas mass at each merger stage.
Such a change may occur when the molecular gas distribution becomes compact 
(the more discussion in the Section \ref{sec:cosize_ms}).
In order for the $\alpha_{\rm CO}$ to account for the observed constancy of molecular gas mass,
$\alpha_{\rm CO}$ would have to change systematically by a factor of 4.5 from the early stage
to the late stage, bringing down the median gas mass in the late stage below the $3\sigma$ spread of the early stage.
Considering the discussion in Section \ref{sec:COandMH2}, this large change in $\alpha_{\rm CO}$ is unlikely.
In addition, even though in the extreme case 
that LIRGs have the Galactic $\alpha_{\rm CO}$ of $4.3 \,M_{\odot}(\rm Kkm\,s^{-1}pc^{-2})^{-1}$ \citep{Bolatto13} 
while ULIRGs have the low $\alpha_{\rm CO}$ of $0.8\,M_{\odot}(\rm Kkm\,s^{-1}pc^{-2})^{-1}$ \citep{DownesSolomon98},
the central molecular gas mass is still constant because of the small number of ULIRGs (eight sources) in our sample.

If the molecular gas mass in the central region through the interaction/merger does not vary, 
then the amount of molecular gas consumed by star formation in the central region are roughly comparable with
the molecular gas mass supplied by the global inflow towards the central region 
during a term between the early stage to the late stage of merger, approximately 1\,Gyr.
In other words, 
the time-averaged SFR over a merger timescale of $\sim 1$\,Gyr, $\langle \rm SFR \rangle$, are comparable with 
the time-averaged rate of gas inflow over the same timescale, $\langle \dot{M}_{\rm inflow} \rangle$.
Thus we can roughly estimate $\langle \dot{M}_{\rm inflow} \rangle$ as 
\begin{equation}
\langle \dot{M}_{\rm inflow} \rangle \approx \langle \rm SFR \rangle \mbox{;  (time-averaged over a merger timescale).}\label{eq:gas_inflow_rate}
\end{equation}
Numerical simulations have predicted two starburst events during a merger \citep[e.g.,][]{MihosHernquist96},
which lasts $\sim 0.3 - 0.5$\,Gyr each \citep{Cox08}.
Since the whole merger event lasts $\sim 1$\,Gyr \citep{Haan11},
we can estimate that the time-averaged SFR is on the order of the median value ($44\,M_{\odot}\rm yr^{-1}$) 
of the galaxies with $\log{(L_{\rm IR}/L_{\odot})} \geq 11.0$ in our sample.
Given large uncertainties,
from Equation \ref{eq:gas_inflow_rate}, 
we estimate $\langle \dot{M}_{\rm inflow} \rangle$ of $\sim 40\,M_{\odot}\rm yr^{-1}$ in our sample.
This rate is significantly larger 
than barred spiral galaxies $> 0.1-1\,M_{\odot}\rm yr^{-1}$ \citep{Sakamoto99}.
This implies that
mergers efficiently transfer the molecular gas towards the central region compared to bars in spiral galaxies.

There are other possible mechanisms involving flows of the molecular gas,  
such as outflows by AGN \citep[e.g.,][]{Garcia14,Sakamoto14,Cicone14}
or by the large-scale wind by starburst \citep[e.g.,][]{Walter02,Cicone14,Cazzoli14},
or the kinetic ejection due to interactions/mergers \citep{Iono04,Kapferer05}.
However, these effects are likely not significant for the mass-loss in our sample.
The kinetic ejections hardly affect depletions of the molecular gas in the central region
because it is effective mainly in the outer disk.
The mass-loss rates by the AGN-driven outflow and the starburst outflow are likely smaller than their SFRs,
because most of our sample is not AGN-dominated galaxies (\citealt{Petric11}; \citetalias{Stierwalt13})
nor shock-heated galaxies \citep{Inami13}.

\begin{figure}
\epsscale{1}
\plotone{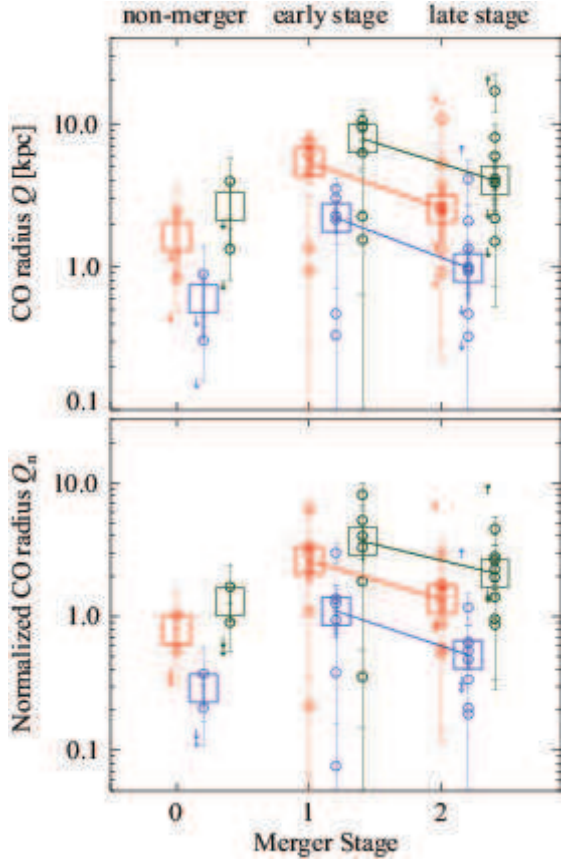}
\caption{
The CO size (a) and the normalized CO size (b) at the each merger stage of the sub-sample.
The CO size and the normalized CO size are colored with the same color scheme as Figure \ref{fig:R_COsize}.
The circles, fine vertical bars, and arrows mean the data values, the errors, and the upper/lower limit values, respectively.
The squares and the vertical bold bars are the median values and its error, respectively.
\label{fig:merger_cosize}}
\end{figure}

\subsubsection{The CO Size and Merger Stage}\label{sec:cosize_ms}
Gas inflow from the outer disk to the central region could downsize the extent of the CO distribution.
We examine the CO size with the merger stage in Figure \ref{fig:merger_cosize}. 
We also show the CO size normalized by a radius in $K_s$-band image, $Q_{\rm n}$, in Figure \ref{fig:merger_cosize}b.
The $K_s$-band radius are the geometric mean of half-light radii between a major axis and a minor axis,
which is taken from the 2MASS Extended Source Catalog \citep{Skrutskie06}.
The seven galaxies with the CO size from the mapping observations are unintentionally early-stage mergers and non-mergers, 
and thus we does not plot these data.

In all three models (Model~A-C), the median values of $Q$ and $Q_{\rm n}$ appear to be decreasing by a factor of $\gtrsim 2$.
However, there are remaining uncertainties: the large dispersion of data, the simplified merger stages, 
and the uncertainties of the estimated CO size, which arise from the assumed distribution profile of CO gas, 
the error in CO intensity, and the pointing error of the observations. 
The decreasing trend, if significant, would be consistent with 
the scenario where the inflow of the molecular gas from the outer disk towards 
the central region in merging LIRGs.

The CO size of the non-mergers are the smallest among the three classes of mergers.
In the molecular gas mass distributions,
the KS test for the non-mergers shows p-values of 0.53 and 0.13 
against the early stage mergers and the late stage mergers, respectively.
It is not rejected that the non-mergers are from a same population 
of the early stage mergers and the late stage mergers.
Therefore, the non-mergers seem to have a centrally concentrated, dense molecular gas.
\citet{Haan11} report that non-mergers in the GOALS galaxies with $L_{\rm IR} \geq 10^{11.4}\,L_{\odot}$
have the most luminous bulge, largest bulge radius, and largest bulge S\'erisic index
compared to other merger stages.
These non-merger LIRGs may be radiating their IR by a different mechanism from the merging LIRGs.

\section{SUMMARY and CONCLUSION}\label{sec:summary}
We present initial results from the CO observation of 79 galaxies in 62 GOALS LIRG systems 
(54 LIRGs and 8 ULIRGs) using the Nobeyama 45\, telescope.
The CO was observed with a single beam towards the central regions 
($r=1.3~\rm kpc - 15~kpc$, $r<5~\rm kpc$ for 80\,\% of our sample) corresponding to 
their brightest MIR positions.
Our CO sample covers the full range of merger stage,
AGN fraction and IR luminosity seen in the GOALS sample.
Using the obtained CO data, we find that

\begin{enumerate}
\item 
CO emission was detected in 68 out of 79 sources,
giving a detection rate of 86\,\%.
Adopting a $\alpha_{\rm CO}$ of $0.6\,M_{\odot}(\rm Kkm\,s^{-1}pc^{-2})^{-1}$,
the molecular gas mass is estimated to be $2.2\times10^8 - 7.0\times10^9\,M_{\odot}$.

\item 
By comparing the NRO45 CO flux to that reported in larger beams in the literature for a subset of 21 LIRGs,
we are able to estimate the extent of the CO emission using a few simple models for the distribution of the molecular gas.
The median values of the CO radii are 2.6\,kpc, 1.0\,kpc, and 4.1\,kpc
in the Gaussian, exponential, and uniform disk models, respectively.
The majority of the galaxies have the CO radii of $\lesssim 4\,\rm kpc$ in  all models.
Despite the low spatial resolutions (FWHM=15\arcsec, 45\arcsec, 55\arcsec) of the single-dish telescopes,
the estimated CO sizes are consistent with interferometric measurements 
for most of the sources with available interferometric data.
We note that our estimation is inaccurate for compact sources with $\mu \lesssim 1\arcsec$.

\item
Any trend is not detected in the molecular gas mass in the central region along the merger sequence. 
Although there is the large dispersion, the median values of the molecular gas mass from the early stage 
to the late stage of merger are constant within their errors.
If the total molecular gas mass decreases along the merger sequence as proposed by \citet{GaoSolomon99},
then this constant central gas mass could imply an inflow towards the central 1--few kpc in these sources.
The CO size appears to be decreasing from the early stage to the late stage of merger by a factor of $\geqq 2$,
although there are large uncertainties remaining.
This might imply a gas inflow from the outer disk towards the central region.
\end{enumerate}

\acknowledgments
We would like to thank the anonymous referee for very useful comments
that helped to further improve this paper.
We also would like to thank the staff at Nobeyama Radio Observatory 
for their support on our observations.
Nobeyama Radio Observatory is a branch of the National Astronomical 
Observatory of Japan, National Institutes of Natural Sciences.
T. Y. acknowledges the financial support from the Global Center 
of Excellence Program by MEXT, Japan through 
the "Nanoscience and Quantum Physics" Project of the Tokyo Institute of Technology.
D.I. is supported by the JSPS KAKENHI Grant Number 15H02074.
This research has made use of the NASA/IPAC Extragalactic Database (NED) 
and the Infrared Science Archive (IRSA) which are operated by the Jet Propulsion Laboratory, 
California Institute of Technology, under contract with the National Aeronautics and Space Administration.

\appendix
\section{Models to Estimate the CO Size}\label{apx:COsize_model}

The CO sizes of 21 sub-sample galaxies are estimated using a combination of flux
from $15\arcsec$, $45\arcsec$, and $55\arcsec$ beam-sized telescopes.
In Equation \ref{eq:Rco}, the ratio of the flux within the $15\arcsec$ beam 
to the flux within $55\arcsec$ or $45\arcsec$ beam is defined.
The numerator is considered to be the total flux of the source.
The denominator corresponds to the flux from the emitting region which is convolved with
the primary beam pattern of the $15\arcsec$ telescope beam.
The primary beam pattern is approximated by a normalized axisymmetric Gaussian distribution 
with the main beam size $\theta_{\rm mb} = 15\arcsec$ and is expressed by the following equation, 
\begin{equation}
P_n(\mbox{\boldmath $r$}) = \exp{(- \frac{4\ln{2}}{\theta_{\rm mb}^2} r^2)}.
\end{equation}
We assume three intensity distribution profiles of the emitting region, 
the Gaussian distribution (Model~A), the azimuthally symmetric exponential distribution (Model~B), 
and the Uniform disk (Model~C).
These are represented as, 
\begin{eqnarray}
I_A(\mbox{\boldmath $r$}) &=& A \exp{(- \frac{4\ln{2}}{\mu_A^2} r^2)}, \\
I_B(\mbox{\boldmath $r$}) &=& A \exp{(- \frac{\ln{2}}{\mu_B} r)}, \\
I_C(\mbox{\boldmath $r$}) &=& A ~~~~~ (r < \mu_C,~{\rm otherwise}~ 0),
\end{eqnarray}
where $\mbox{\boldmath $r$}$ is the radius from the peak of the intensity, and $A$ is a constant. 
The $\mu_A$, $\mu_B$, and $\mu_C$ are the CO sizes for each model. 
For Model~A and B, the CO sizes $\mu_A$ and $\mu_B$ are identical to the HWHMs.
The $\mu_C$ is the radius of the disk in Model~C. 
As long as the center of the beam coincides with the center of the intensity distribution,
the flux ratio in Equation \ref{eq:Rco} is rewritten as, 
\begin{equation}\label{eq:Rco_apx}
R_{\rm CO} = \frac{\int_0^{2\pi} \int_0^{\infty} I_X(\mbox{\boldmath $r$},\mu) P_n(\mbox{\boldmath $r$})\,r\, drd\phi}
{\int_0^{2\pi} \int_0^{\infty} I_X(\mbox{\boldmath $r$},\mu)\,r\, drd\phi}
\end{equation}
where $X$ is replaced by indices of Model~A, B or C.
By solving Equations \ref{eq:Rco_apx} for $\mu$, we obtain the CO size.
We can easily isolate $\mu_A$ as follows,
\begin{equation}
\mu_A = 0.5 \sqrt{\frac{1}{R_{\rm CO}}-1}\, .
\end{equation}
Model~B and C cannot be solved analytically for $\mu_B$ and $\mu_C$. 
We solve Model B and C numerically.
The equations to be solved are,
\begin{eqnarray}
&&{\rm Model~B:~} R_{\rm CO} = \frac{(\ln{2})^2}{\mu_B^2} \exp{\left(\frac{\theta_{\rm mb}\ln{2}}{16\mu_B^2}\right)} \int_0^{\infty} \exp{\left\{- \frac{4\ln{2}}{\theta_{\rm mb}^2} \left(r + \frac{\theta_{\rm mb}^2}{8\mu_B}\right)^2 \right\}}\,r\,dr \label{eq:Rco_apx_b}\\
&&{\rm Model~C:~} R_{\rm CO} = \frac{\theta_{\rm mb}^2}{4\mu_C^2\ln{2}} \left\{ 1 - \exp{\left( - \frac{2\mu_C^2\ln{2}}{\theta_{\rm mb}^2}\right)} \right\} .
\end{eqnarray}
The integration in Equation \ref{eq:Rco_apx_b} was performed by using the IDL function \verb|QROMO|.
The solutions of the CO sizes for given flux ratios are displayed in Figure \ref{fig:R_COsize}.
The accuracy of this estimate and the difference between the models are discussed in Section \ref{sec:co_estimate}.

\section{The Measurement of the CO Size from the Mapping Observations}\label{apx:mapping}
We also conducted the half-beam spacing mapping for 31 LIRGs during the 2012 and 2013 semesters.
The basic reductions were done in a similar manner to the pointing observations, and the maps are produced from 2--35 positions.
For seven galaxies, the CO size are measured. They have CO detections of more than three positions along a axis and 
positions of their peak intensities does not deviate from the galactic center.

\begin{figure}
\epsscale{1.}
\plotone{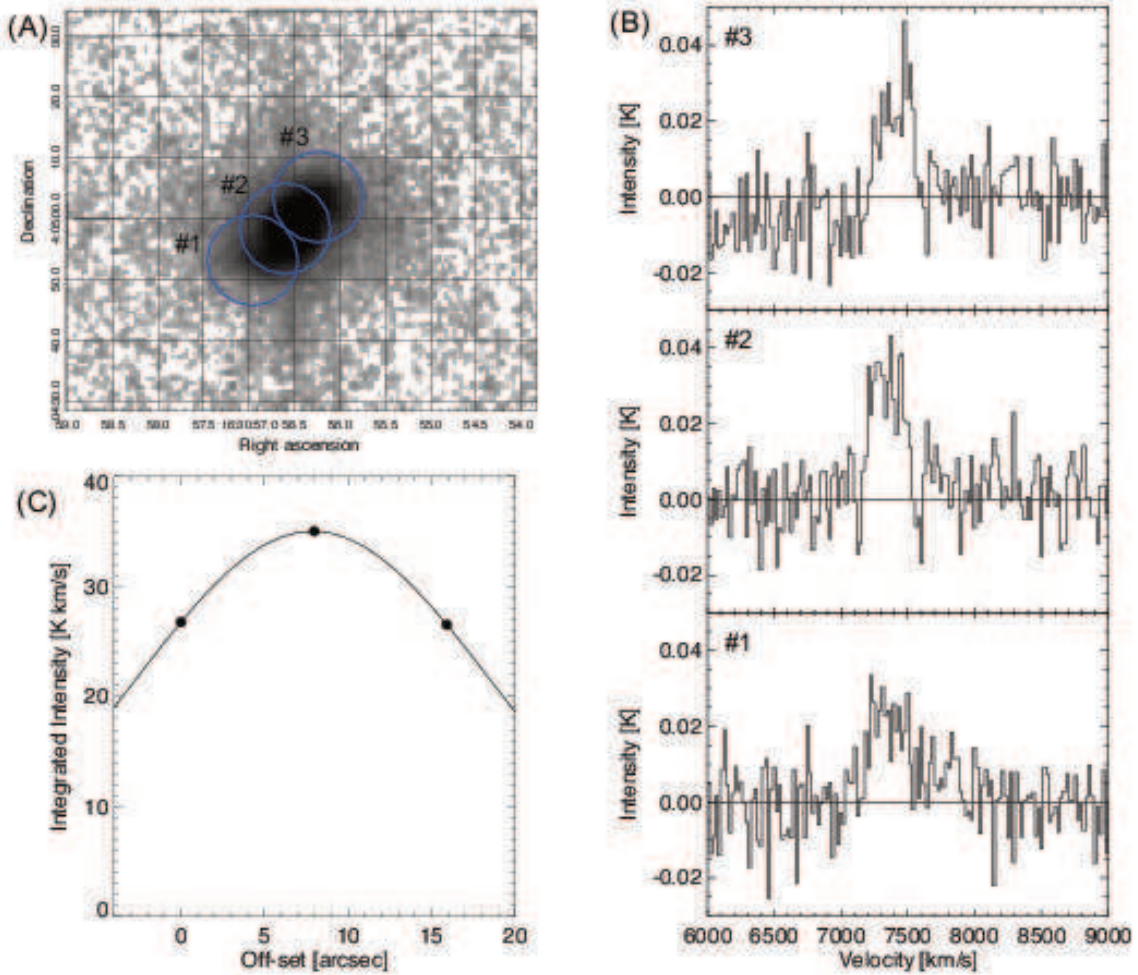}
\caption{
An example of the mapping observation.
The object is CGCG~052-037.
(A): The beam positions (blue circles) of the NRO45 mapping on the {\it Spitzer}/IRAC $8\,\mu m$ image (gray scale).
(B): The spectra of CO emission from the three positions.
(C): The radial profile of the integrated intensity.  
The $x$-axis indicates the off-set from the position \#1. The solid line is the fitted Gaussian function.
\label{fig:map}}
\end{figure}

A Gaussian function is fitted to the radial profile of integrated intensities along a axis.
The obtained Gaussian HWHM is deconvolved by the NRO45 beam profile function 
which is approximated by a Gaussian with $\theta_{\rm mb}$ of $15\arcsec$.
The deconvolved HWHM is defined as the CO size of the mapping observations.
When the Gaussian HWHM is smaller than the beam size, the Gaussian HWHM is the upper limit of the CO size.
An example of the mapping observation and a radial profile are shown in Figure \ref{fig:map}.

\section{Notes on Individual Sources}\label{apx:note_obj}
We describe the characteristics of four sources
which have excessively large or small flux ratio ($R_{\rm CO} \gtrsim 1.9$ or $R_{\rm CO} \lesssim 0.2$),
and one source which may have CO intensity distributions that possibly deviate from our assumed models.

{\it NGC 1275} --- 
This object is a cD galaxy at the center of the Perseus cluster.
A strong radio continuum comes from Perseus A (3C~084) located in the center of NGC~1275.
The cold molecular gas is associated with filament structures
seen in H$\alpha$ \citep{Salome06,Salome08a,Salome08b}.
The NRO45 observation of the $15\arcsec$ beam shows that 
this source has 98.2\,Jy\,km\,s$^{-1}$, which is 75\,\% larger than 
35.7\,Jy\,km\,s$^{-1}$ given in \citet{Young95} by the $45\arcsec$ beam,
and therefore has a high $R_{\rm CO}$ of 2.75 (see Section \ref{sec:comparison_flux}).
The flux towards the center of NGC~1275 reported in the literature vary greatly.
For example,
\citet{Lazareff89} reported a flux of $123\pm4$\,Jy\,km\,s$^{-1}$ with the $21\arcsec$ beam at IRAM~30\,m.
\citet{Mirabel89} obtained a flux of 65.8\,Jy\,km\,s$^{-1}$ with the the $55\arcsec$ beam at NRAO~12\,m,
and \citet{Evans05} measured the flux to be $104\pm1$\,Jy\,km\,s$^{-1}$ with $55\arcsec$ beam 
at the Kitt Peak 12\,m Telescope.
An observation by \citet{Salome08b} reported the central flux of NGC~1275 
to be $26.7\pm5.8$\,Jy\,km\,s$^{-1}$ with IRAM~30\,m.
This undetermined flux may be due to instabilities of the baseline in the spectra due to the strong radio continuum.

We note that $\Delta V_{\rm FWHM}$ may be overestimated because of the low S/N ratio.
In the velocity range with the high S/N ratio the $\Delta V_{\rm FWHM}$ is estimated to be 240\,km\,s$^{-1}$.
\citet{Lazareff89}, \citet{Mirabel89}, \citet{Evans05}, and \citet{Salome08b} estimate $\Delta V_{\rm FWHM}$
to be 200 -- 300\,km\,s$^{-1}$.

{\it UGC 05101} ---
This ULIRG shows a high flux ratio of $R_{\rm CO} = 1.87$.
The flux is $141.4\pm9.7$\,Jy\,km\,s$^{-1}$ based on the NRO45 observation, 
while 75.5\,Jy\,km\,s$^{-1}$ based on the IRAM 30\,m observation \citep{GaoSolomon04a}.
\citet{Solomon97} obtained 77.2\,Jy\,km\,s$^{-1}$ with the IRAM 30\,m.
The velocity widths of \citet{GaoSolomon04a} and \citet{Solomon97} are consistent with each other, as well as the flux.
Their spectra are located in the heliocentric velocity range from 11500\,km\,s$^{-1}$ to 12000\,km\,s$^{-1}$, 
Our spectrum shows an additional redshifted component +500 \,km\,s$^{-1}$ 
and consequently the broad velocity width of 840\,km\,s$^{-1}$.
Due to this redshifted feature, our measurement of the flux is larger than those 
by \citet{GaoSolomon04a} and \citet{Solomon97}.

{\it Mrk 231} ---
For this ULIRG, the flux measurements by NRO45 and \citetalias{Sanders91} show
$110.8\pm5.0$\,Jy\,km\,s$^{-1}$ and 56.0\,Jy\,km\,s$^{-1}$, respectively.
Therefore Mrk~231 has a high flux ratio of 1.98.
Our measurement is consistent with the IRAM 30\,m observation with $22\arcsec$ beam, 
which gives $S_{\rm CO}\Delta V = 109$\,Jy\,km\,s$^{-1}$ \citep{Solomon97}.
Additionally the CO spectrum by \citetalias{Sanders91} shows two peaks, 
while the spectrum of our observation and \citet{Solomon97} show a single peak.
Thus the measurement by \citetalias{Sanders91} may be doubtful.
The S/N ratio of the observation by \citetalias{Sanders91} appears to be lower
compared to our observation and the observation by \citet{Solomon97}.

{\it MCG+04-48-002S (NGC~6921, MCG~+04-48-001)} ---
This is a less-IR luminous object ($L_{\rm IR} = 10^{10.68}\,L_{\odot}$), 
accompanying a luminous galaxy MCG~+04-48-002N (MCG~+04-48-002, 
$L_{\rm IR} = 10^{11.06}\,L_{\odot}$) offset on the sky by 1.5$\arcmin$.
Our flux measurement shows $37.1\pm5.8$\,Jy\,km\,s$^{-1}$.
\citetalias{GaoSolomon04a} obtain a flux of 188\,Jy\,km\,s$^{-1}$ with a $22\arcsec$ beam observation.
Thus a flux ratio $R_{\rm CO}$ is as low as 0.2.
\citet{Young86,Young95} reported a larger flux of 312\,Jy\,km\,s$^{-1}$.
Both observations of \citetalias{GaoSolomon04a} and \citet{Young86,Young95}, however, 
suffer from baseline uncertainties due to relatively narrow bandwidths.
Moreover, the line profiles are quite different between \citetalias{GaoSolomon04a} and \citet{Young86}.
The flux from \citetalias{GaoSolomon04a} is consistent with a flux estimated 
from the $M_{\rm H_2}$ -- $L_{\rm IR}$ relation for the starburst galaxies in \citet{Daddi10b}.

{\it NGC 7771S1} ---
This object is an edge-on galaxy companying two galaxies 
NGC~7771N (NGC~7769) and NGC~7771S2 (NGC~7770).
The 24\,$\mu$m peak is located at the galactic center, 
while the two additional components of 24\,$\mu$m are seen 
at both edges of the galaxy (see Figure \ref{fig:spectra}).
Interferometric observation of the galaxy shows an elongated CO distribution 
towards the east edge \citep{Dale05}.
Therefore the CO spatial structure might not match any of models to estimate
the CO extent in Section \ref{sec:co_estimate}.

The line profile shows three distinct peaks (see Figure \ref{fig:spectra}).
This three-peak profile is also seen in the 14\,m observation (\citetalias{Sanders91}; \citealt{Young95}).
The detection by the smaller $15\arcsec$ beam observation indicates
that the origin of the profile is within the central 2.2\,kpc.
There may be a rotating molecular ring around the galactic nucleus which produces the secondary peaks,
and concentrated molecular clouds which emerge as the central peak in the spectrum.
The rotating molecular ring might be directly associated with the starburst ring existing around the nucleus \citep{Smith99}.

{\it Mrk 331} ---
This object is a disk galaxy at the pre-merger stage (\citealt{Haan11}; \citetalias{Stierwalt13}).
The CO we obtain shows a much lower flux ($45.0\pm8.8$\,Jy\,km\,s$^{-1}$)
than the 12\,m observation \citepalias[346.4\,Jy\,km\,s$^{-1}$,][]{GaoSolomon04a}.
The flux ratio of 0.13 is the lowest in our sample.
Other flux measurements of Mrk~331 report 371\,Jy\,km\,s$^{-1}$ with the IRAM 30\,m \citep{Solomon97} 
and 392\,Jy\,km\,s$^{-1}$ with the FCRAO 14\,m \citep{Young95}.
The CO gas in this object could be extremely extended.

\clearpage

\begin{figure*}[t]
\includegraphics[width=\textwidth]{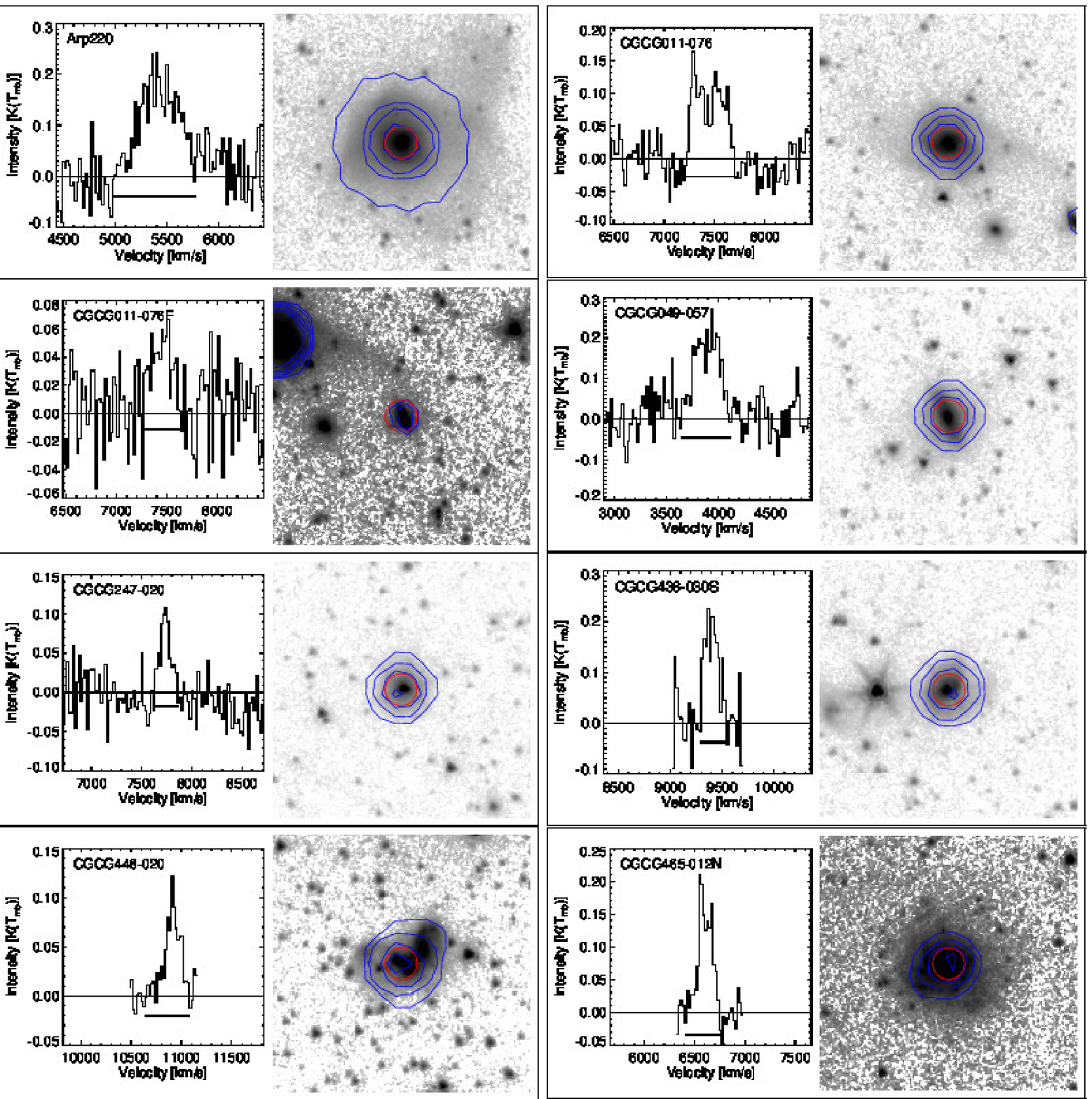}
\caption{The CO spectrum and the $3.6\,\mu\rm m$ image ({\sl Spitzer}/IRAC) for the sample. 
The intensity is in the main beam temperature, $T_{\rm mb}$.
The velocity range on x-axis is 2000\,km\,s$^{-1}$.
The solid bar under the spectrum represents the integral range to derive the integral intensity.
For each source, we show the $15''$ diameter NRO45 CO beam (red) and 
the contours of the {\sl Spitzer}/MIPS $24\,\mu\rm m$ (blue contours) on the {\sl Spitzer}/IRAC image (grey scale). 
The contour levels are 0.2, 0.4, 0.6, and 0.8 of the peak intensity in logarithmic scale.
The image covers an area of $2{'} \times 2{'}$ on the sky.   
\label{fig:spectra}}
\end{figure*}

\begin{figure*} 
\figurenum{\ref{fig:spectra}}
\includegraphics[width=\textwidth]{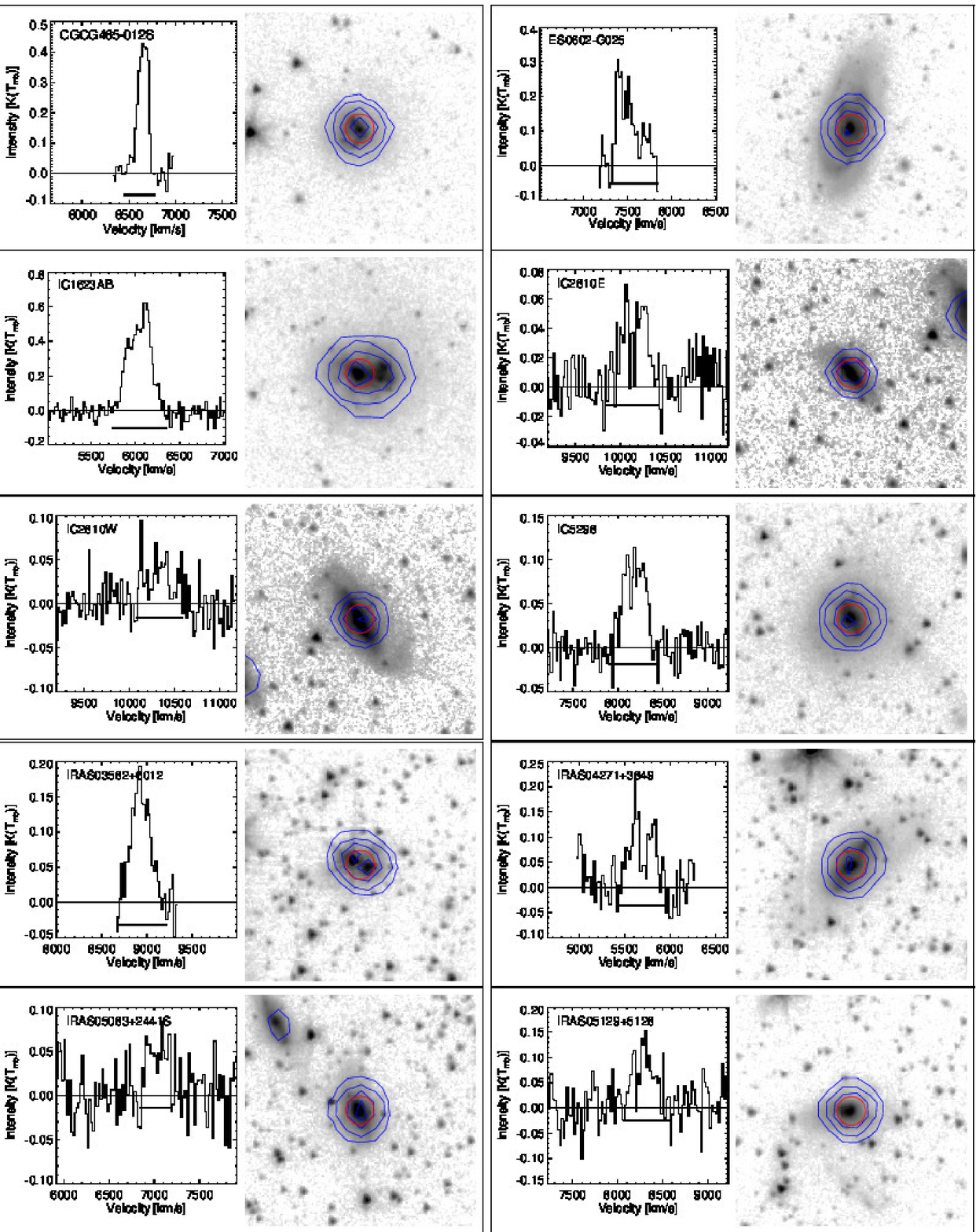}
\caption{ {\it Continued.}
}
\end{figure*}

\begin{figure*} 
\figurenum{\ref{fig:spectra}}
\includegraphics[width=\textwidth]{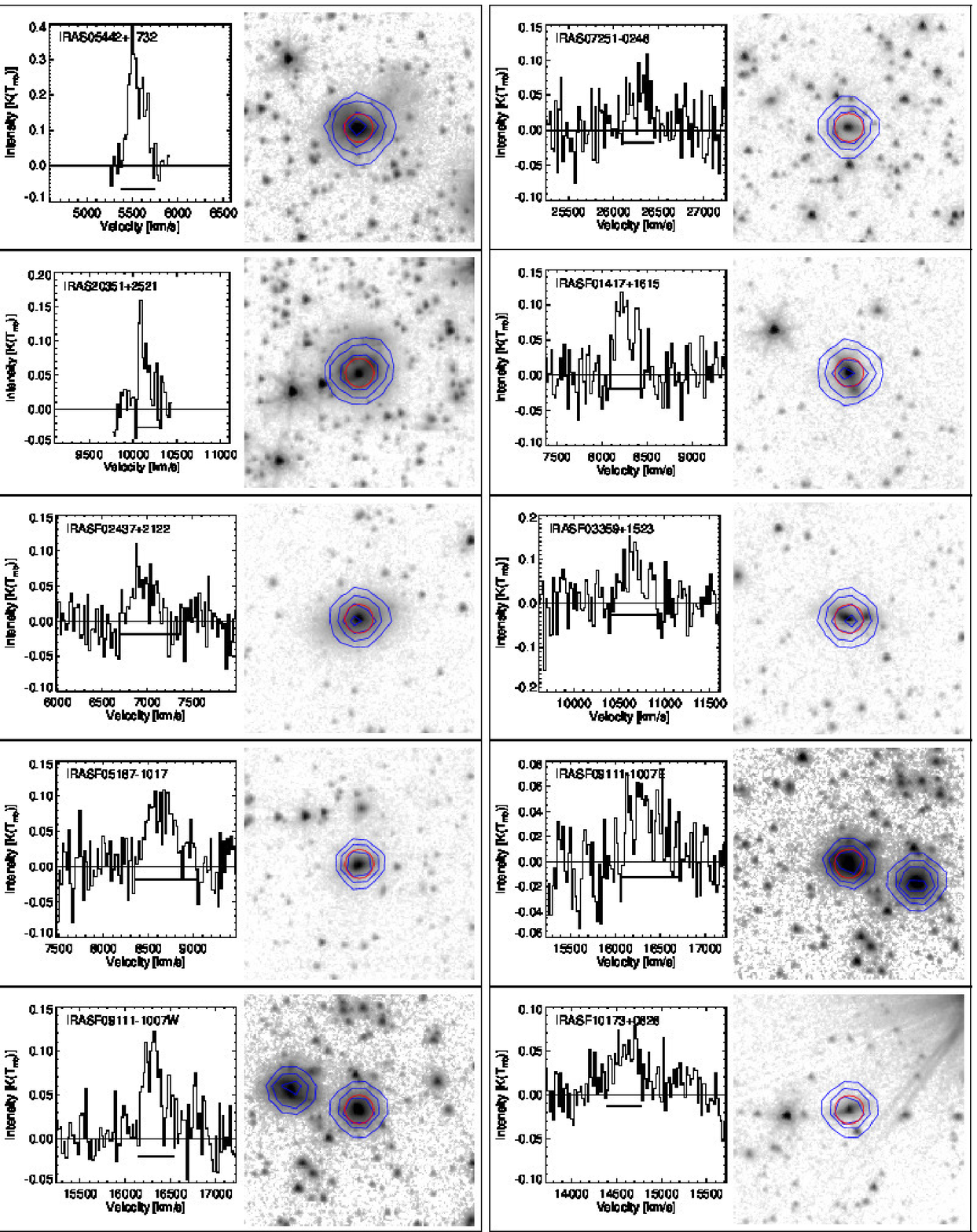}
\caption{ {\it Continued.}
}
\end{figure*}

\begin{figure*} 
\figurenum{\ref{fig:spectra}}
\includegraphics[width=\textwidth]{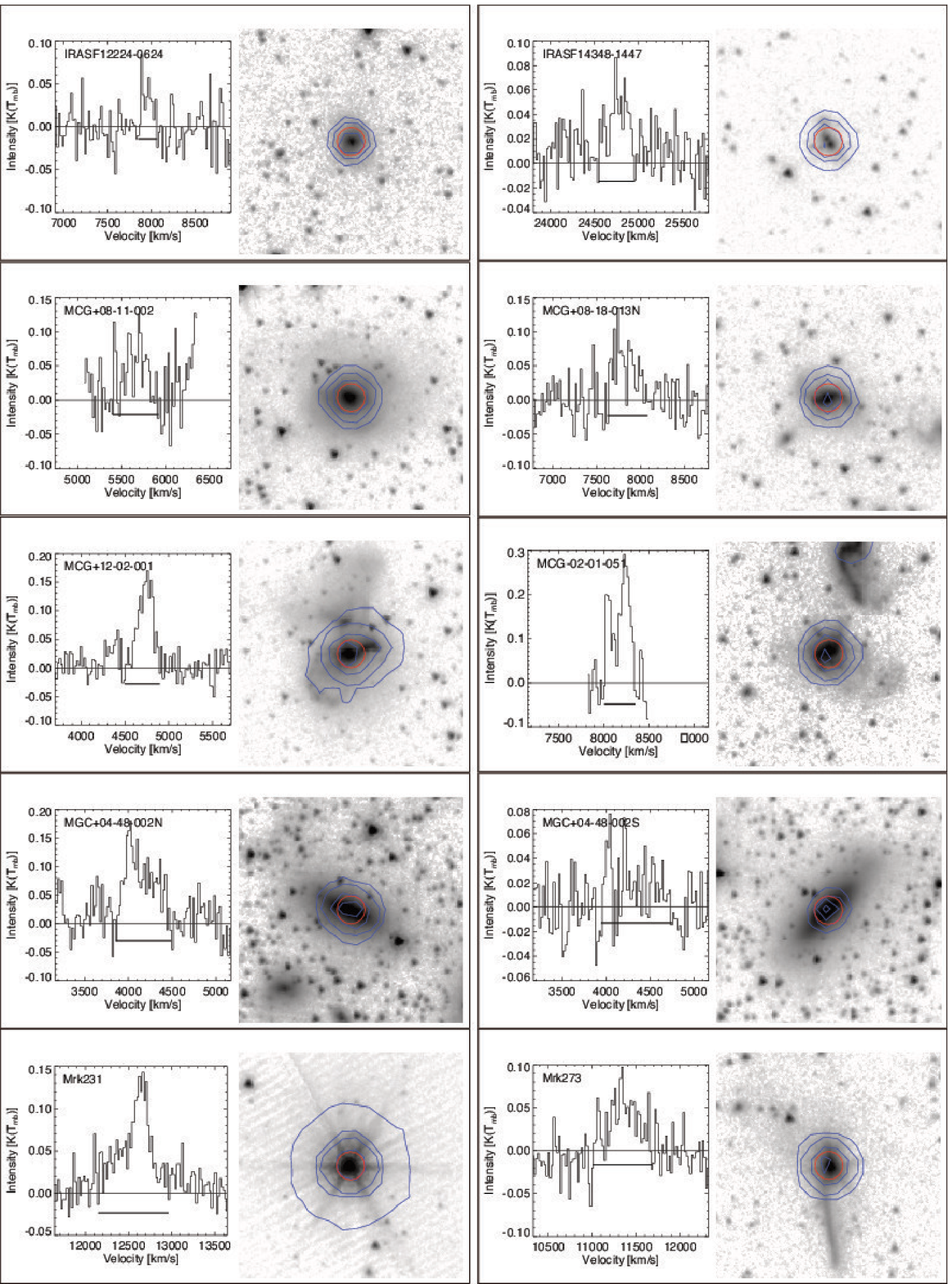}
\caption{ {\it Continued.}
}
\end{figure*}

\begin{figure*} 
\figurenum{\ref{fig:spectra}}
\includegraphics[width=\textwidth]{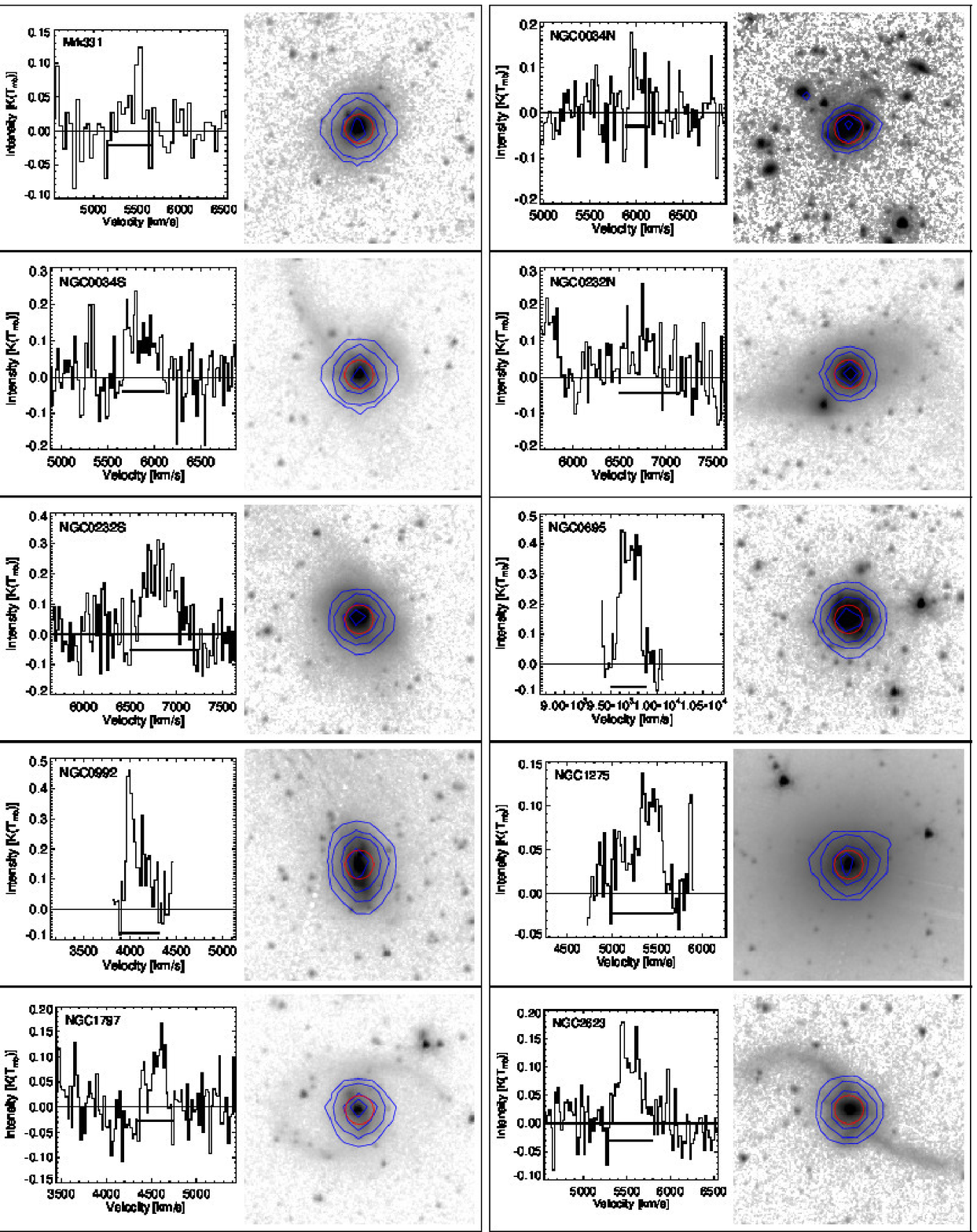}
\caption{ {\it Continued,}
but the $4.5\,\mu\rm m$ image ({\sl Spitzer}/IRAC) of NGC~0034N is used because the $3.6\,\mu\rm m$ image is not available.
}
\end{figure*}

\begin{figure*} 
\figurenum{\ref{fig:spectra}}
\includegraphics[width=\textwidth]{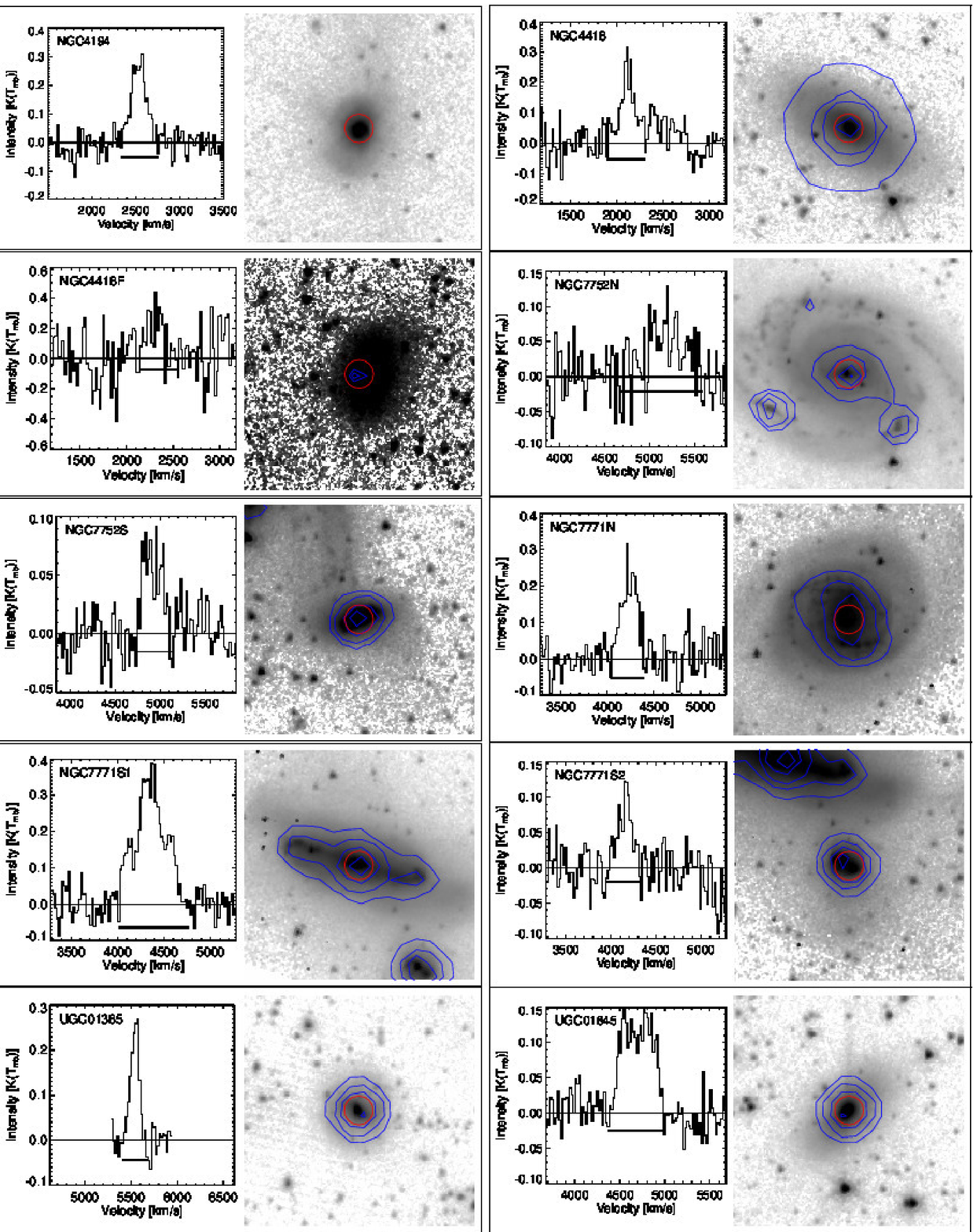}
\caption{ {\it Continued,}
but {\sl Spitzer}/MIPS $24\,\mu\rm m$ image of NGC~4194 is not available.
}
\end{figure*}

\begin{figure*} 
\figurenum{\ref{fig:spectra}}
\includegraphics[width=\textwidth]{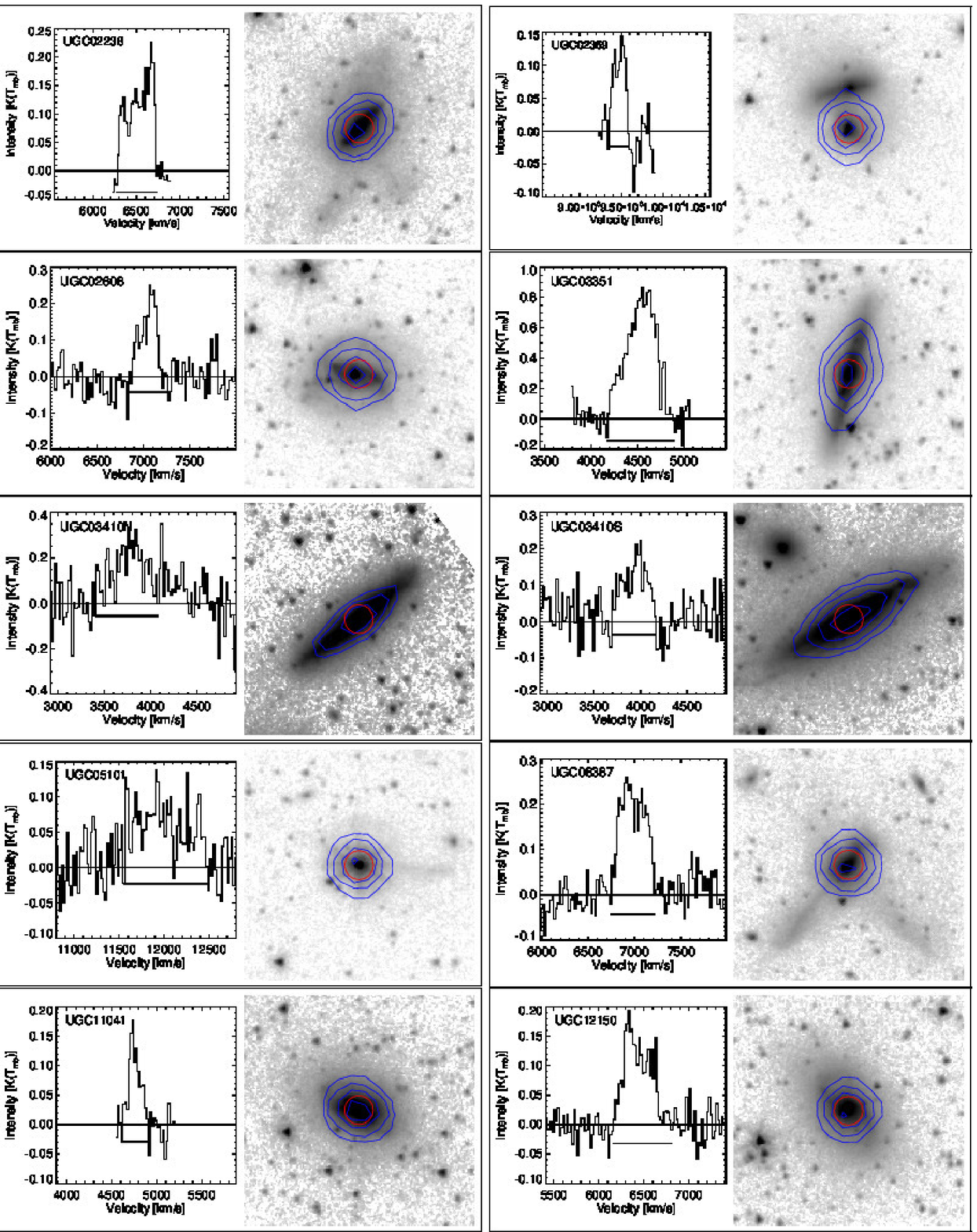}
\caption{ {\it Continued.}
}
\end{figure*}

{\renewcommand\arraystretch{0.3}
\begin{deluxetable*}{lllrrr}
\tabletypesize{\tiny}
\tablewidth{0pt}
\tablecaption{GOALS sources observed with Nobeyama 45\,m\label{tab:source}}
\tablehead{
\colhead{Galaxies}     & \colhead{R.A.(J2000)}   & \colhead{Dec.(J2000)}   &    \colhead{Velocity}         &  \colhead{$D_{L}$}   & \colhead{$\log(L_{\rm IR}/L_{\odot})$}\\
\colhead{}                   & \colhead{}                      & \colhead{}                       &    \colhead{(km\,s$^{-1}$)} &  \colhead{(Mpc)}       &      \\ 
\colhead{(1)}               & \colhead{(2)}                 & \colhead{(3)}                   &    \colhead{(4)}                 &  \colhead{(5)}           &      \colhead{(6)}
}
\startdata
NGC 0034S  & 		 $00^{\rm h}11^{\rm m}06.55^{\rm s}$  &  $-12^{\circ}06^{\prime}27.8^{\prime\prime}$ 	 & 5881 & 84.1 & 11.34\tablenotemark{H} \\
NGC 0034N  & 		 $00^{\rm h}11^{\rm m}10.42^{\rm s}$  &  $-12^{\circ}01^{\prime}16.1^{\prime\prime}$ 	 & 5881 & 84.1 & 10.57\tablenotemark{H} \\
MCG -02-01-051  &  	 $00^{\rm h}18^{\rm m}50.86^{\rm s}$  &  $-10^{\circ}22^{\prime}37.5^{\prime\prime}$ 	 & 8159 & 117.5 & 11.48 \\
NGC 0232S  &  		 $00^{\rm h}42^{\rm m}45.82^{\rm s}$  &  $-23^{\circ}33^{\prime}41.7^{\prime\prime}$ 	 & 6647 & 95.2 & 11.28\tablenotemark{D} \\
NGC 0232N  & 		 $00^{\rm h}42^{\rm m}52.82^{\rm s}$  &  $-23^{\circ}32^{\prime}28.5^{\prime\prime}$ 	 & 6647 & 95.2 & 10.93\tablenotemark{D} \\
MCG +12-02-001  & 	 $00^{\rm h}54^{\rm m}03.88^{\rm s}$  &  $+73^{\circ}05^{\prime}05.6^{\prime\prime}$ 	 & 4706 & 69.8 & 11.50 \\
IC 1623AB  & 		 $01^{\rm h}07^{\rm m}47.43^{\rm s}$  &  $-17^{\circ}30^{\prime}25.1^{\prime\prime}$ 	 & 6016 & 85.5 & 11.71 \\
MCG -03-04-014  & 	 $01^{\rm h}10^{\rm m}08.93^{\rm s}$  &  $-16^{\circ}51^{\prime}11.1^{\prime\prime}$ 	 & 10040 & 144 & 11.65 \\
CGCG 436-030  & 	         $01^{\rm h}20^{\rm m}02.59^{\rm s}$  &  $+14^{\circ}21^{\prime}42.5^{\prime\prime}$ 	 & 9362 & 134 & 11.69 \\
IRAS F01417+1651  &        $01^{\rm h}44^{\rm m}30.53^{\rm s}$  &  $+17^{\circ}06^{\prime}08.9^{\prime\prime}$ 	 & 8375 & 119 & 11.64 \\
NGC 0695  & 		 $01^{\rm h}51^{\rm m}14.29^{\rm s}$  &  $+22^{\circ}34^{\prime}55.2^{\prime\prime}$ 	 & 9735 & 139 & 11.68 \\
UGC 01385  & 		 $01^{\rm h}54^{\rm m}53.76^{\rm s}$  &  $+36^{\circ}55^{\prime}04.5^{\prime\prime}$ 	 & 5621 & 79.8 & 11.05 \\
UGC 01845  & 		 $02^{\rm h}24^{\rm m}07.89^{\rm s}$  &  $+47^{\circ}58^{\prime}11.3^{\prime\prime}$ 	 & 4679 & 67 & 11.12 \\
NGC 0992  & 		 $02^{\rm h}37^{\rm m}25.50^{\rm s}$  &  $+21^{\circ}06^{\prime}03.9^{\prime\prime}$ 	 & 4141 & 58 & 11.07 \\
UGC 02238  & 		 $02^{\rm h}46^{\rm m}17.50^{\rm s}$  &  $+13^{\circ}05^{\prime}44.9^{\prime\prime}$ 	 & 6560 & 92.4 & 11.33 \\
IRAS F02437+2122  &        $02^{\rm h}46^{\rm m}39.13^{\rm s}$  &  $+21^{\circ}35^{\prime}10.5^{\prime\prime}$ 	 & 6987 & 98.8 & 11.16 \\
UGC 02369  & 		 $02^{\rm h}54^{\rm m}01.81^{\rm s}$  &  $+14^{\circ}58^{\prime}14.3^{\prime\prime}$ 	 & 9558 & 136 & 11.67 \\
UGC 02608  & 		 $03^{\rm h}15^{\rm m}01.25^{\rm s}$  &  $+42^{\circ}02^{\prime}09.2^{\prime\prime}$ 	 & 6998 & 100 & 11.41 \\
NGC 1275  & 		 $03^{\rm h}19^{\rm m}48.18^{\rm s}$  &  $+41^{\circ}30^{\prime}42.2^{\prime\prime}$ 	 & 5264 & 75 & 11.26 \\
IRAS F03359+1523  &        $03^{\rm h}38^{\rm m}47.02^{\rm s}$  &  $+15^{\circ}32^{\prime}53.1^{\prime\prime}$ 	 & 10613 & 152 & 11.55 \\
CGCG 465-012N  & 	         $03^{\rm h}54^{\rm m}07.64^{\rm s}$  &  $+15^{\circ}59^{\prime}24.7^{\prime\prime}$ 	 & 6662 & 94.3 & 10.54\tablenotemark{D} \\
CGCG 465-012S  & 	         $03^{\rm h}54^{\rm m}15.97^{\rm s}$  &  $+15^{\circ}55^{\prime}43.8^{\prime\prime}$ 	 & 6662 & 94.3 & 11.09\tablenotemark{D} \\
IRAS 03582+6012  & 	 $04^{\rm h}02^{\rm m}32.55^{\rm s}$  &  $+60^{\circ}20^{\prime}39.7^{\prime\prime}$ 	 & 8997 & 131 & 11.43 \\
IRAS 04271+3849  & 	 $04^{\rm h}30^{\rm m}33.10^{\rm s}$  &  $+38^{\circ}55^{\prime}48.4^{\prime\prime}$ 	 & 5640 & 80.8 & 11.11 \\
NGC 1797F  & 		 $05^{\rm h}07^{\rm m}44.56^{\rm s}$  &  $-07^{\circ}58^{\prime}09.8^{\prime\prime}$ 	 & 4441 & 63.4 & 9.20\tablenotemark{X} \\
NGC 1797  & 		 $05^{\rm h}07^{\rm m}44.82^{\rm s}$  &  $-08^{\circ}01^{\prime}08.6^{\prime\prime}$ 	 & 4441 & 63.4 & 11.03\tablenotemark{X} \\
CGCG 468-002S  & 	         $05^{\rm h}08^{\rm m}19.67^{\rm s}$  &  $+17^{\circ}21^{\prime}47.7^{\prime\prime}$ 	 & 5454 & 77.9 & 10.92\tablenotemark{D} \\
CGCG 468-002N  & 	         $05^{\rm h}08^{\rm m}21.19^{\rm s}$  &  $+17^{\circ}22^{\prime}08.2^{\prime\prime}$ 	 & 5454 & 77.9 & 10.92\tablenotemark{D} \\
IRAS 05083+2441  &         $05^{\rm h}11^{\rm m}25.83^{\rm s}$  &  $+24^{\circ}45^{\prime}18.7^{\prime\prime}$ 	 & 6915 & 99.2 & 11.26 \\
IRAS 05129+5128  & 	 $05^{\rm h}16^{\rm m}55.94^{\rm s}$  &  $+51^{\circ}31^{\prime}57.0^{\prime\prime}$ 	 & 8224 & 120 & 11.42 \\
IRAS F05187-1017  & 	 $05^{\rm h}21^{\rm m}06.52^{\rm s}$  &  $-10^{\circ}14^{\prime}45.6^{\prime\prime}$ 	 & 8474 & 122 & 11.3 \\
IRAS 05223+1908  & 	 $05^{\rm h}25^{\rm m}16.65^{\rm s}$  &  $+19^{\circ}10^{\prime}48.5^{\prime\prime}$ 	 & 8867 & 128 & 11.65 \\
MCG +08-11-002  & 	 $05^{\rm h}40^{\rm m}43.66^{\rm s}$  &  $+49^{\circ}41^{\prime}41.8^{\prime\prime}$ 	 & 5743 & 83.7 & 11.46 \\
UGC 03351  & 	 	 $05^{\rm h}45^{\rm m}48.01^{\rm s}$  &  $+58^{\circ}42^{\prime}03.7^{\prime\prime}$ 	 & 4455 & 65.8 & 11.28 \\
IRAS 05442+1732  & 	 $05^{\rm h}47^{\rm m}11.16^{\rm s}$  &  $+17^{\circ}33^{\prime}47.2^{\prime\prime}$ 	 & 5582 & 80.5 & 11.30 \\
UGC 03410N  & 		 $06^{\rm h}13^{\rm m}58.14^{\rm s}$  &  $+80^{\circ}28^{\prime}34.5^{\prime\prime}$ 	 & 3921 & 59.7 & 10.29\tablenotemark{D} \\
UGC 03410S  & 		 $06^{\rm h}14^{\rm m}29.64^{\rm s}$  &  $+80^{\circ}26^{\prime}59.4^{\prime\prime}$ 	 & 3921 & 59.7 & 11.03\tablenotemark{D} \\
IRAS 07251-0248  & 	 $07^{\rm h}27^{\rm m}37.53^{\rm s}$  &  $-02^{\circ}54^{\prime}54.2^{\prime\prime}$ 	 & 26249 & 400 & 12.39 \\
NGC 2623  & 		 $08^{\rm h}38^{\rm m}24.14^{\rm s}$  &  $+25^{\circ}45^{\prime}16.7^{\prime\prime}$ 	 & 5549 & 84.1 & 11.60 \\
IRAS F09111-1007W  &       $09^{\rm h}13^{\rm m}36.42^{\rm s}$  &  $-10^{\circ}19^{\prime}29.8^{\prime\prime}$ 	 & 16231 & 246 & 11.86\tablenotemark{D} \\
IRAS F09111-1007E  &       $09^{\rm h}13^{\rm m}38.82^{\rm s}$  &  $-10^{\circ}19^{\prime}20.0^{\prime\prime}$ 	 & 16231 & 246 & 11.62\tablenotemark{D} \\
UGC 05101  & 		 $09^{\rm h}35^{\rm m}51.66^{\rm s}$  &  $+61^{\circ}21^{\prime}11.5^{\prime\prime}$ 	 & 11802 & 177 & 12.01 \\
MCG +08-18-013S  & 	 $09^{\rm h}36^{\rm m}30.87^{\rm s}$  &  $+48^{\circ}28^{\prime}10.1^{\prime\prime}$ 	 & 7777 & 117 & 9.93\tablenotemark{H} \\
MCG +08-18-013N  & 	 $09^{\rm h}36^{\rm m}37.16^{\rm s}$  &  $+48^{\circ}28^{\prime}28.2^{\prime\prime}$ 	 & 7777 & 117 & 11.32\tablenotemark{H} \\
IRAS F10173+0828  &        $10^{\rm h}20^{\rm m}00.19^{\rm s}$  &  $+08^{\circ}13^{\prime}33.9^{\prime\prime}$ 	 & 14716 & 224 & 11.86\tablenotemark{X} \\
IRAS F10173+0828F  &       $10^{\rm h}20^{\rm m}01.41^{\rm s}$  &  $+08^{\circ}11^{\prime}31.7^{\prime\prime}$ 	 & 14716 & 224 & 9.60\tablenotemark{X} \\
CGCG 011-076F  & 	         $11^{\rm h}21^{\rm m}08.34^{\rm s}$  &  $-02^{\circ}59^{\prime}38.0^{\prime\prime}$ 	 & 7464 & 117 & 10.02\tablenotemark{H} \\
CGCG 011-076  & 	         $11^{\rm h}21^{\rm m}12.24^{\rm s}$  &  $-02^{\circ}59^{\prime}01.9^{\prime\prime}$ 	 & 7464 & 117 & 11.41\tablenotemark{H} \\
IC 2810W  & 		 $11^{\rm h}25^{\rm m}45.00^{\rm s}$  &  $+14^{\circ}40^{\prime}36.4^{\prime\prime}$ 	 & 10192 & 157 & 11.45\tablenotemark{H} \\
IC 2810E  & 		 $11^{\rm h}25^{\rm m}49.48^{\rm s}$  &  $+14^{\circ}40^{\prime}06.7^{\prime\prime}$ 	 & 10192 & 157 & 11.20\tablenotemark{H} \\
NGC 4194  & 		 $12^{\rm h}14^{\rm m}09.64^{\rm s}$  &  $+54^{\circ}31^{\prime}36.1^{\prime\prime}$ 	 & 2501 & 43 & 11.10 \\
IRAS F12224-0624  & 	 $12^{\rm h}25^{\rm m}03.95^{\rm s}$  &  $-06^{\circ}40^{\prime}52.9^{\prime\prime}$ 	 & 7902 & 125 & 11.36 \\
NGC 4418  & 		 $12^{\rm h}26^{\rm m}54.66^{\rm s}$  &  $-00^{\circ}52^{\prime}39.5^{\prime\prime}$ 	 & 2179 & 36.5 & 11.19\tablenotemark{X} \\
NGC 4418F  & 		 $12^{\rm h}27^{\rm m}04.96^{\rm s}$  &  $-00^{\circ}54^{\prime}25.8^{\prime\prime}$ 	 & 2179 & 36.5 & 8.59\tablenotemark{X} \\
Mrk 231  &                 $12^{\rm h}56^{\rm m}14.10^{\rm s}$  &  $+56^{\circ}52^{\prime}25.7^{\prime\prime}$ 	 & 12642 & 192 & 12.57 \\
UGC 08387  & 		 $13^{\rm h}20^{\rm m}35.31^{\rm s}$  &  $+34^{\circ}08^{\prime}22.7^{\prime\prime}$ 	 & 6985 & 110 & 11.73 \\
Mrk 273  & 		 $13^{\rm h}44^{\rm m}42.14^{\rm s}$  &  $+55^{\circ}53^{\prime}13.9^{\prime\prime}$ 	 & 11326 & 173 & 12.21 \\
CGCG 247-020  & 	         $14^{\rm h}19^{\rm m}43.35^{\rm s}$  &  $+49^{\circ}14^{\prime}11.5^{\prime\prime}$ 	 & 7716 & 120 & 11.39 \\
IRAS F14348-1447  & 	 $14^{\rm h}37^{\rm m}38.32^{\rm s}$  &  $-15^{\circ}00^{\prime}22.7^{\prime\prime}$ 	 & 24802 & 387 & 12.39 \\
CGCG 049-057  & 	         $15^{\rm h}13^{\rm m}13.07^{\rm s}$  &  $+07^{\circ}13^{\prime}32.3^{\prime\prime}$ 	 & 3897 & 65.4 & 11.35 \\
Arp 220  & 		 $15^{\rm h}34^{\rm m}57.25^{\rm s}$  &  $+23^{\circ}30^{\prime}11.1^{\prime\prime}$ 	 & 5434 & 87.9 & 12.28 \\
IRAS F17207-0014  & 	 $17^{\rm h}23^{\rm m}21.98^{\rm s}$  &  $-00^{\circ}17^{\prime}00.6^{\prime\prime}$ 	 & 12834 & 198 & 12.46 \\
UGC 11041  & 		 $17^{\rm h}54^{\rm m}51.82^{\rm s}$  &  $+34^{\circ}46^{\prime}34.3^{\prime\prime}$ 	 & 4881 & 77.5 & 11.11 \\
CGCG 141-034  & 	         $17^{\rm h}56^{\rm m}56.61^{\rm s}$  &  $+24^{\circ}01^{\prime}02.0^{\prime\prime}$ 	 & 5944 & 93.4 & 11.20 \\
CGCG 142-034S  & 	         $18^{\rm h}16^{\rm m}33.84^{\rm s}$  &  $+22^{\circ}06^{\prime}38.4^{\prime\prime}$ 	 & 5599 & 88.1 & 10.64\tablenotemark{D} \\
CGCG 142-034N  & 	         $18^{\rm h}16^{\rm m}40.69^{\rm s}$  &  $+22^{\circ}06^{\prime}46.2^{\prime\prime}$ 	 & 5599 & 88.1 & 11.03\tablenotemark{D} \\
MCG +04-48-002S  & 	 $20^{\rm h}28^{\rm m}28.85^{\rm s}$  &  $+25^{\circ}43^{\prime}24.6^{\prime\prime}$ 	 & 4167 & 64.2 & 10.68\tablenotemark{X} \\
MCG +04-48-002N  & 	 $20^{\rm h}28^{\rm m}35.03^{\rm s}$  &  $+25^{\circ}44^{\prime}00.6^{\prime\prime}$ 	 & 4167 & 64.2 & 11.06\tablenotemark{X} \\
IRAS 20351+2521  & 	 $20^{\rm h}37^{\rm m}17.72^{\rm s}$  &  $+25^{\circ}31^{\prime}38.0^{\prime\prime}$ 	 & 10102 & 151 & 11.61 \\
CGCG 448-020  & 	         $20^{\rm h}57^{\rm m}24.33^{\rm s}$  &  $+17^{\circ}07^{\prime}38.3^{\prime\prime}$ 	 & 10822 & 161 & 11.94 \\
ESO 602-G025  & 	         $22^{\rm h}31^{\rm m}25.44^{\rm s}$  &  $-19^{\circ}02^{\prime}03.9^{\prime\prime}$ 	 & 7507 & 110 & 11.34 \\
UGC 12150  & 		 $22^{\rm h}41^{\rm m}12.20^{\rm s}$  &  $+34^{\circ}14^{\prime}56.2^{\prime\prime}$ 	 & 6413 & 93.5 & 11.35 \\
IC 5298  & 		 $23^{\rm h}16^{\rm m}00.65^{\rm s}$  &  $+25^{\circ}33^{\prime}23.7^{\prime\prime}$ 	 & 8221 & 119 & 11.60 \\
NGC 7752S  & 		 $23^{\rm h}46^{\rm m}58.48^{\rm s}$  &  $+29^{\circ}27^{\prime}31.8^{\prime\prime}$ 	 & 5120 & 73.6 & 11.07\tablenotemark{D} \\
NGC 7752N  & 		 $23^{\rm h}47^{\rm m}04.74^{\rm s}$  &  $+29^{\circ}29^{\prime}00.2^{\prime\prime}$ 	 & 5120 & 73.6 & 11.07\tablenotemark{D} \\
NGC 7771N  & 		 $23^{\rm h}51^{\rm m}03.90^{\rm s}$  &  $+20^{\circ}09^{\prime}00.8^{\prime\prime}$ 	 & 4277 & 61.2 & 10.74\tablenotemark{D} \\
NGC 7771S2  & 		 $23^{\rm h}51^{\rm m}22.43^{\rm s}$  &  $+20^{\circ}05^{\prime}46.9^{\prime\prime}$ 	 & 4277 & 61.2 & 10.67\tablenotemark{D} \\
NGC 7771S1  & 		 $23^{\rm h}51^{\rm m}24.79^{\rm s}$  &  $+20^{\circ}06^{\prime}41.7^{\prime\prime}$ 	 & 4277 & 61.2 & 11.17\tablenotemark{D} \\
Mrk 331  & 		 $23^{\rm h}51^{\rm m}26.72^{\rm s}$  &  $+20^{\circ}35^{\prime}09.5^{\prime\prime}$ 	 & 5541 & 79.3 & 11.50 \\
\enddata
\tablecomments{
Col. (1): The galaxy name. The interacting galaxies are labeled by ``N", ``E", ``S", and ``W" in pair galaxies.
Col. (2), (3): The observing coordinates, which is the brightest position in $24\,\mu\rm m$ image from MIPS/{\it Spitzer}.
Col. (4), (5), and (6): The heliocentric velocity of galaxies, the luminosity distance, and the IR luminosity taken from  \citet{Armus09}, who calculate the IR luminosity using four the IRAS bands from 12\,$\mu$m to 100\,$\mu$m and the derivation in \citet{SandersMirabel96}.
For resolved galaxies, the individual IR luminosities are shown, 
which are taken from \citet{Howell10} (H) and \citet{Diaz-Santos10} (D), or this work (X) (see text for details).
}
\end{deluxetable*}
}

{\renewcommand\arraystretch{0.5}
\begin{deluxetable*}{lcrrrrrrr}
\tabletypesize{\tiny}
\tablewidth{0pt}
\tablecaption{Observational results\label{tab:COresult}}
\tablehead{
\colhead{Galaxies}         &  \colhead{Backend} &
\colhead{$\Delta V_0$}  &  \colhead{$\Delta V_{\rm FWHM}$} &
\colhead{$I_{\rm CO}$}  &  
\colhead{$S_{\rm CO}\Delta V$}  &  \colhead{$S_{\rm CO}\Delta V$(ref)} &
\colhead{$L^{\prime}_{\rm CO}$} &\colhead{$M_{\rm H_2}$} \\
\colhead{}                       & \colhead{}              &
\colhead{(km\,s$^{-1}$)}   & \colhead{(km\,s$^{-1}$)}                 &
\colhead{(K\,km\,s$^{-1}$)} & 
\colhead{(Jy\,km\,s$^{-1}$)}            &\colhead{(Jy\,km\,s$^{-1}$)}                          &
\colhead{($10^{8}\,L^{\prime}$)}  &\colhead{($10^{8}\,M_{\odot}$)} \\ 
\colhead{(1)}                   &\colhead{(2)}           & 
\colhead{(3)}                   &    \colhead{(4)}                              &  
\colhead{(5)}                   &
\colhead{(6)}                   &\colhead{(7)}                                              &
\colhead{(8)}                                 &\colhead{(9)}
}
\startdata
NGC0034S         &  SAM45  &  420  &  293  &  $46.13\pm 5.67$  &  $113.0\pm 13.9$  &     & 	 $18.4\pm 2.27$  &  $11.1\pm 1.36$ \\
NGC0034N         &  SAM45  &  200  &  166  &  $12.86\pm 3.63$  &  $31.50\pm 8.89$  &     & 	 $5.14\pm 1.45$  &  $3.08\pm 0.871$ \\
MCG-02-01-051    &  AOS    &  320  &  277  &  $51.58\pm 2.43$  &  $126.4\pm 6.0$   &     & 	 $39.4\pm 1.86$  &  $23.6\pm 1.11$ \\
NGC0232S         &  SAM45  &  680  &  436  &  $89.24\pm 8.17$  &  $218.6\pm 20.0$  &     & 	 $45.4\pm 4.15$  &  $27.2\pm 2.49$ \\
NGC0232N         &  SAM45  &  600  &  373  &  $38.57\pm 5.97$  &  $94.50\pm 14.64$ &     & 	 $19.6\pm 3.04$  &  $11.8\pm 1.82$ \\
MCG+12-02-001 	 &  SAM45  &  395  &  178  &  $30.79\pm 1.97$  &  $75.44\pm 4.84$  &     & 	 $8.58\pm 0.55$  &  $5.15\pm 0.33$ \\
IC1623AB    	 &  SAM45  &  580  &  307  &  $170.6\pm 4.1$   &  $418.1\pm 10.0$  &  493.5\tablenotemark{c} & 	 $70.4\pm 1.68$  &  $42.3\pm 1.01$ \\
MCG-03-04-014 	 &  AOS    &       &       &  $<29.33$ 		   &  $<71.86$ 		 &     & 	  $<33$  &   $<19.8$ \\
CGCG436-030 	 &  AOS    &  240  &  206  &  $32.42\pm 1.92$  &  $79.43\pm 4.69$  &     & 	 $31.8\pm 1.88$  &  $19.1\pm 1.13$ \\
IRASF01417+1651  &  SAM45  &  340  &  272  &  $23.44\pm 2.50$  &  $57.44\pm 6.11$  &  63.0\tablenotemark{c} & 	 $18.3\pm 1.95$  &  $11\pm 1.17$ \\
NGC0695 	 &  AOS    &  360  &  250  &  $96.53\pm 2.80$  &  $236.5\pm 6.9$ 	 &  199.9\tablenotemark{b} & 	 $102\pm 2.95$  &  $60.9\pm 1.77$ \\
UGC01385 	 &  AOS    &  240  &  103  &  $28.59\pm 0.80$  &  $70.04\pm 1.95$  &     & 	 $10.3\pm 0.287$  &  $6.19\pm 0.172$ \\
UGC01845 	 &  SAM45  &  600  &  458  &  $55.98\pm 1.80$  &  $137.2\pm 4.4$ 	 &     & 	 $14.4\pm 0.461$  &  $8.62\pm 0.277$ \\
NGC0992 	 &  AOS    &  400  &  187  &  $71.47\pm 2.73$  &  $175.1\pm 6.7$ 	 &  207.9\tablenotemark{c} & 	 $13.8\pm 0.527$  &  $8.3\pm 0.316$ \\
UGC02238 	 &  AOS    &  440  &  382  &  $53.22\pm 0.94$  &  $130.4\pm 2.3$ 	 &  210.0\tablenotemark{c} & 	 $25.5\pm 0.452$  &  $15.3\pm 0.271$ \\
IRASF02437+2122  &  SAM45  &  640  &  235  &  $16.64\pm 2.72$  &  $40.76\pm 6.67$  &     & 	 $9.08\pm 1.49$  &  $5.45\pm 0.892$ \\
UGC02369 	 &  AOS    &  220  &  173  &  $20.55\pm 0.91$  &  $50.34\pm 2.23$  &     & 	 $20.7\pm 0.917$  &  $12.4\pm 0.55$ \\
UGC02608 	 &  SAM45  &  420  &  223  &  $36.30\pm 3.59$  &  $88.94\pm 8.80$  &     & 	 $20.3\pm 2.01$  &  $12.2\pm 1.21$ \\
NGC1275*	 &  AC45   &  680  &  538  &  $40.08\pm 1.83$  &  $98.21\pm 4.48$  &  35.7\tablenotemark{b} & 	 $12.8\pm 0.585$  &  $7.69\pm 0.351$ \\
IRASF03359+1523  &  SAM45  &  500  &  318  &  $28.36\pm 3.97$  &  $69.48\pm 9.73$  &  133.0\tablenotemark{c} & 	 $35.4\pm 4.95$  &  $21.2\pm 2.97$ \\
CGCG465-012N 	 &  AOS    &  320  &  148  &  $28.29\pm 1.01$  &  $69.32\pm 2.48$  &     & 	 $14.1\pm 0.504$  &  $8.47\pm 0.302$ \\
CGCG465-012S 	 &  AOS    &  300  &  136  &  $60.63\pm 1.18$  &  $148.6\pm 2.9$ 	 &     & 	 $30.2\pm 0.588$  &  $18.1\pm 0.353$ \\
IRAS03582+6012 	 &  AOS    &  520  &  211  &  $40.76\pm 1.01$  &  $99.86\pm 2.48$  &     & 	 $38.4\pm 0.951$  &  $23\pm 0.571$ \\
IRAS04271+3849 	 &  AC45   &  520  &  255  &  $35.37\pm 2.39$  &  $86.65\pm 5.85$  &     & 	 $13.1\pm 0.883$  &  $7.85\pm 0.53$ \\
NGC1797F 	 &  SAM45  &       &       &  $<26.66$ 		   &  $<65.31$ 		 &     & 	  $<6.14$  &   $<3.69$ \\
NGC1797 	 &  SAM45  &  370  &  257  &  $23.37\pm 3.17$  &  $57.27\pm 7.78$  &     & 	 $5.39\pm 0.732$  &  $3.23\pm 0.439$ \\
CGCG468-002S 	 &  SAM45  &       &       &  $<24.53$ 		   &  $<60.10$ 		 &     & 	  $<8.45$  &   $<5.07$ \\
CGCG468-002N 	 &  AC45   &       &       &  $<24.33$ 		   &  $<59.60$ 		 &     & 	  $<8.38$  &   $<5.03$ \\
IRAS05083+2441S  &  SAM45  &  320  &  389  &  $14.17\pm 2.03$  &  $34.72\pm 4.97$  &     & 	 $7.8\pm 1.12$  &  $4.68\pm 0.67$ \\
IRAS05129+5128 	 &  SAM45  &  300  &  361  &  $31.05\pm 2.90$  &  $76.08\pm 7.09$  &     & 	 $24.7\pm 2.3$  &  $14.8\pm 1.38$ \\
IRASF05187-1017  &  SAM45  &  685  &  515  &  $34.64\pm 3.22$  &  $84.87\pm 7.89$  &     & 	 $28.4\pm 2.64$  &  $17.1\pm 1.59$ \\
IRAS05223+1908 	 &  SAM45  &       &       &  $<20.50$ 		   &  $<50.22$ 		 &     & 	  $<18.4$  &   $<11.1$ \\
MCG+08-11-002 	 &  AC45   &  500  &  378  &  $22.57\pm 2.36$  &  $55.29\pm 5.78$  &     & 	 $8.95\pm 0.935$  &  $5.37\pm 0.561$ \\
UGC03351 	 &  AC45   &  700  &  334  &  $310.6\pm 5.7$   &  $760.9\pm 14.0$  &     & 	 $77.1\pm 1.42$  &  $46.3\pm 0.853$ \\
IRAS05442+1732 	 &  AOS    &  340  &  226  &  $61.16\pm 2.06$  &  $149.84\pm 5.05$ &     & 	 $22.5\pm 0.758$  &  $13.5\pm 0.455$ \\
UGC03410N 	 &  SAM45  &  665  &  586  &  $108.3\pm 9.4$   &  $265.2\pm 22.9$  &     & 	 $22.2\pm 1.92$  &  $13.3\pm 1.15$ \\
UGC03410S 	 &  SAM45  &  480  &  344  &  $54.64\pm 5.87$  &  $133.9\pm 14.4$  &     & 	 $11.2\pm 1.21$  &  $6.73\pm 0.724$ \\
IRAS07251-0248 	 &  SAM45  &  340  &  293  &  $15.56\pm 2.00$  &  $38.12\pm 4.89$  &     & 	 $116\pm 14.9$  &  $69.6\pm 8.92$ \\
NGC2623 	 &  SAM45  &  500  &  230  &  $39.21\pm 2.86$  &  $96.07\pm 7.00$  &  161.3\tablenotemark{b} & 	 $15.7\pm 1.15$  &  $9.44\pm 0.687$ \\
IRASF09111-1007W &  SAM45  &  390  &  303  &  $25.01\pm 1.65$  &  $61.28\pm 4.03$ &     & 	 $77.4\pm 5.1$  &  $46.5\pm 3.06$ \\
IRASF09111-1007E &  SAM45  &  635  &  596  &  $20.88\pm 1.99$  &  $51.15\pm 4.87$ &     & 	 $64.6\pm 6.15$  &  $38.8\pm 3.69$ \\
UGC05101*	 &  SAM45  &  930  &  839  &  $57.71\pm 3.94$  &  $141.4\pm 9.7$   &  75.5\tablenotemark{a} & 	 $96.5\pm 6.59$  &  $57.9\pm 3.95$ \\
MCG+08-18-013S 	 &  SAM45  &       &       &  $<22.84$ 		   &  $<55.95$ 		 &     & 	  $<17.3$  &   $<10.4$ \\
MCG+08-18-013N 	 &  SAM45  &  180  &  216  &  $22.56\pm 1.59$  &  $55.26\pm 3.89$  &     & 	 $17.1\pm 1.21$  &  $10.3\pm 0.724$ \\
IRASF10173+0828  &  SAM45  &  390  &  346  &  $16.66\pm 1.70$  &  $40.83\pm 4.16$  &  63.0\tablenotemark{c} & 	 $43.4\pm 4.42$  &  $26\pm 2.65$ \\
IRASF10173+0828F &  SAM45  &       &       &  $<17.22$ 		   &  $<42.20$ 		 &     & 	  $<44.9$  &   $<26.9$ \\
CGCG011-076F 	 &  SAM45  &  390  &  286  &  $13.56\pm 1.83$  &  $33.23\pm 4.47$  &     & 	 $10.3\pm 1.39$  &  $6.2\pm 0.835$ \\
CGCG011-076 	 &  SAM45  &  480  &  389  &  $39.84\pm 2.00$  &  $97.62\pm 4.90$  &     & 	 $30.4\pm 1.53$  &  $18.2\pm 0.915$ \\
IC2810W 	 &  SAM45  &  505  &  460  &  $15.88\pm 1.96$  &  $38.91\pm 4.81$  &  101.5\tablenotemark{c} & 	 $21.2\pm 2.62$  &  $12.7\pm 1.57$ \\
IC2810E 	 &  SAM45  &  605  &  354  &  $17.00\pm 1.65$  &  $41.65\pm 4.03$  &     & 	 $22.7\pm 2.2$  &  $13.6\pm 1.32$ \\
NGC4194 	 &  SAM45  &  410  &  184  &  $59.72\pm 3.11$  &  $146.3\pm 7.6$   &  143.5\tablenotemark{c} & 	 $6.45\pm 0.336$  &  $3.87\pm 0.202$ \\
IRASF12224-0624  &  SAM45  &  210  &  142  &  $6.76\pm 1.38$   &  $16.57\pm 3.39$  &     & 	 $5.86\pm 1.2$  &  $3.51\pm 0.719$ \\
NGC4418 	 &  SAM45  &  410  &  120  &  $48.39\pm 3.66$  &  $118.6\pm 9.0$   &  164.5\tablenotemark{c} & 	 $3.78\pm 0.286$  &  $2.27\pm 0.171$ \\
NGC4418F 	 &  SAM45  &  400  &  210  &  $60.08\pm 12.66$ &  $147.2\pm 31.0$  &     & 	 $4.69\pm 0.989$  &  $2.82\pm 0.593$ \\
Mrk231*		 &  SAM45  &  815  &  194  &  $45.24\pm 2.05$  &  $110.8\pm 5.0$   &  56.0\tablenotemark{c} & 	 $88.3\pm 4$  &  $53\pm 2.4$ \\
UGC08387 	 &  SAM45  &  475  &  369  &  $79.32\pm 2.66$  &  $194.3\pm 6.5$   &  177.2\tablenotemark{b} & 	 $53.7\pm 1.8$  &  $32.2\pm 1.08$ \\
Mrk273 		 &  SAM45  &  675  &  591  &  $24.46\pm 2.32$  &  $59.94\pm 5.68$  &  80.5\tablenotemark{c} & 	 $39.3\pm 3.72$  &  $23.6\pm 2.23$ \\
CGCG247-020 	 &  SAM45  &  235  &  88   &  $12.66\pm 1.69$  &  $31.03\pm 4.13$   &     & 	 $10.1\pm 1.35$  &  $6.08\pm 0.809$ \\
IRASF14348-1447  &  SAM45  &  410  &  280  &  $15.29\pm 1.46$  &  $37.45\pm 3.58$  &  59.5\tablenotemark{c} & 	 $108\pm 10.3$  &  $64.9\pm 6.19$ \\
CGCG049-057 	 &  SAM45  &  465  &  288  &  $61.34\pm 4.92$  &  $150.3\pm 12.0$  &  119.0\tablenotemark{c} & 	 $15.1\pm 1.21$  &  $9.07\pm 0.727$ \\
Arp220 		 &  SAM45  &  760  &  468  &  $84.64\pm 4.21$  &  $207.4\pm 10.3$  &  329.0\tablenotemark{c} & 	 $37.1\pm 1.85$  &  $22.3\pm 1.11$ \\
IRASF17207-0014  &  SAM45  &       &       &  $<19.18$ 		   &  $<46.99$ 		 &  212.7\tablenotemark{a} & 	  $<39.7$  &   $<23.8$ \\
UGC11041 	 &  AOS    &  280  &  106  &  $20.49\pm 0.97$  &  $50.21\pm 2.38$  &     & 	 $7.03\pm 0.334$  &  $4.22\pm 0.2$ \\
CGCG141-034 	 &  AOS    &       &       &  $<11.20$ 		   &  $<27.43$ 		 &     & 	  $<5.52$  &   $<3.31$ \\
CGCG142-034S 	 &  AOS    &       &       &  $<9.85$ 		   &  $<24.13$ 		 &     & 	  $<4.33$  &   $<2.6$ \\
CGCG142-034N 	 &  AC45   &       &       &  $<14.53$ 		   &  $<35.61$ 		 &     & 	  $<6.39$  &   $<3.84$ \\
MCG+04-48-002S*	 &  SAM45  &  780  &  651  &  $15.12\pm 2.38$  &  $37.06\pm 5.84$  &  187.5\tablenotemark{a} & 	 $3.58\pm 0.564$  &  $2.15\pm 0.339$ \\
MCG+04-48-002N 	 &  SAM45  &  600  &  402  &  $48.02\pm 3.63$  &  $117.7\pm 8.9$   &     & 	 $11.4\pm 0.86$  &  $6.83\pm 0.516$ \\
IRAS20351+2521 	 &  AOS    &  250  &  96   &  $15.98\pm 0.87$  &  $39.15\pm 2.13$  &     & 	 $19.8\pm 1.07$  &  $11.9\pm 0.645$ \\
CGCG448-020 	 &  AOS    &  440  &  104  &  $16.33\pm 0.90$  &  $40.00\pm 2.21$  &     & 	 $22.8\pm 1.26$  &  $13.7\pm 0.757$ \\
ESO602-G025 	 &  AOS    &  500  &  180  &  $61.89\pm 2.51$  &  $151.64\pm 6.14$ &     & 	 $41.7\pm 1.69$  &  $25\pm 1.01$ \\
UGC12150	 &  SAM45  &  650  &  364  &  $55.53\pm 2.48$  &  $136.06\pm 6.07$ &     & 	 $27.3\pm 1.22$  &  $16.4\pm 0.731$ \\
IC5298 		 &  SAM45  &  519  &  281  &  $26.01\pm 1.89$  &  $63.72\pm 4.63$  &  84.0\tablenotemark{c} & 	 $20.3\pm 1.48$  &  $12.2\pm 0.886$ \\
NGC7752S 	 &  SAM45  &  360  &  260  &  $17.46\pm 1.42$  &  $42.77\pm 3.49$  &     & 	 $5.39\pm 0.44$  &  $3.23\pm 0.264$ \\
NGC7752N 	 &  SAM45  &  860  &  563  &  $32.14\pm 3.74$  &  $78.74\pm 9.17$  &     & 	 $9.92\pm 1.15$  &  $5.95\pm 0.693$ \\
NGC7771N 	 &  SAM45  &  340  &  136  &  $47.64\pm 3.18$  &  $116.7\pm 7.8$   &     & 	 $10.2\pm 0.683$  &  $6.15\pm 0.41$ \\
NGC7771S2 	 &  SAM45  &  320  &  153  &  $18.91\pm 1.70$  &  $46.32\pm 4.17$  &     & 	 $4.07\pm 0.366$  &  $2.44\pm 0.22$ \\
NGC7771S1 	 &  SAM45  &  720  &  239  &  $132.0\pm 3.4$  &  $323.3\pm 8.4$    &  380.5\tablenotemark{b} & 	 $28.4\pm 0.74$  &  $17\pm 0.444$ \\
Mrk331* 	 &  SAM45  &  475  &  80   &  $18.36\pm 3.58$  &  $44.99\pm 8.78$   &  346.4\tablenotemark{a} & 	 $6.55\pm 1.28$  &  $3.93\pm 0.767$ \\
\enddata
\tablecomments{
Col. (1): Galaxy name. 
Col. (2): Spectrometer used in our CO observation.
Col. (3): Full velocity width at zero intensity of the CO emission line.
Col. (4): Full velocity width at half maximum of the CO line.
Col. (5): CO intensity on the temperature scale of $T_{\rm mb}$.
Col. (6): CO flux. A conversion factor from K to Jy is 2.45\,JyK($T_{\rm mb}$)$^{-1}$.
Col. (7): CO flux in the literature (a = \citetalias{GaoSolomon04a}, b = \citet{Young95}, and c = \citetalias{Sanders91}).
Conversion factors from K to Jy are 4.95\,JyK($T_{\rm mb}$)$^{-1}$, 42\,JyK($T_A^*$)$^{-1}$, and 35\,JyK($T_R^*$)$^{-1}$ for the IRAM 30\,m telescope,
FCRAO 14\,m telescope \citep{Young95}, and NRAO 12\,m telescope (the NRAO 12\,m User's Manual 1990 edition, Fig. 14.), respectively.
For 14\,m observations in the $T_R^*$ scale, 3.15\,JyK($T_R^*$) is used assuming $\eta_{\rm fss} = 0.75$.
Col. (8): CO luminosity,
Col. (9): Molecular gas mass calculated with the $0.6\,M_{\odot}(\rm Kkm\,s^{-1}pc^{-2})^{-1}$ \citep{Papadopoulos12}.
The asterisk denotes the objects with the additional information about the uncertain CO flux or the line profile in Appendix \ref{apx:note_obj}.
}
\end{deluxetable*}
}

\clearpage
\begin{turnpage}
\begin{deluxetable}{lllllllllll}
\tablecolumns{10}
\tabletypesize{\scriptsize}
\tablewidth{0pt}
\tablecaption{The estimated CO size\label{tab:COsize}}
\tablehead{
\colhead{Galaxies}         &  \colhead{$R_{\rm CO}$} &
\multicolumn{2}{c}{A: Gaussian}   & \colhead{}  &
\multicolumn{2}{c}{B: Exponential}   &  \colhead{}  &
\multicolumn{2}{c}{C: Uniform disk}  &
\colhead{Literature} \\
\cline{3-4} \cline{6-7} \cline{9-10}
\colhead{}                       & \colhead{}   &
\colhead{$\mu_A$ ($^{\prime\prime}$)}  & \colhead{$Q_A$ (kpc)}  &  \colhead{}  &
\colhead{$\mu_B$ ($^{\prime\prime}$)}  & \colhead{$Q_B$ (kpc)}  &  \colhead{}  &
\colhead{$\mu_C$ ($^{\prime\prime}$)}  & \colhead{$Q_C$ (kpc)}  &
\colhead{$\mu$ ($^{\prime\prime}$)} \\
\colhead{(1)}                   &\colhead{(2)}          & 
\colhead{(3)}                   &\colhead{(4)}          &  \colhead{}  &
\colhead{(5)}                   &\colhead{(6)}          & \colhead{}  &
\colhead{(7)}                   &\colhead{(8)}          & \colhead{(9)}
}
\startdata
IC1623AB        & 0.847  & $ 3.18 \pm 2.48 $ & $ 1.32 \pm 1.03: $ & & $ 1.13 \pm 1.61: $ & $ 0.47 \pm 0.67 $ & & $ 5.26 \pm 3.53 $ & $ 2.18 \pm 1.46 $ & 4\tablenotemark{a,1}, 7.5$\times$4\tablenotemark{d,2}\\
IRASF01417+1651 & 0.912  & $ 2.34 \pm 3.15: $ & $ 1.35 \pm 1.82: $ & & $ 0.81 \pm 2.8: $  & $ 0.47 \pm 1.61: $ & & $ 3.9 \pm 4.05: $  & $ 2.25 \pm 2.34: $ \\
NGC0695         &$<0.939$& $<1.91$           & $<1.29$           & & $<0.67$           & $<0.45          $ & & $<3.21$           & $<2.16$ \\
NGC0992         & 0.842  & $ 3.24 \pm 2.45 $ & $ 0.91 \pm 0.69 $ & & $ 1.15 \pm 1.53: $ & $ 0.32 \pm 0.43: $ & & $ 5.36 \pm 3.49 $ & $ 1.51 \pm 0.98 $ \\
UGC02238        & 0.621  & $ 5.86 \pm 1.85 $ & $ 2.63 \pm 0.83 $ & & $ 2.23 \pm 0.75 $ & $ 1.00 \pm 0.33 $ & & $ 9.20 \pm 2.70 $ & $ 4.12 \pm 1.21 $ & 6.0\tablenotemark{j,3}\\
IRASF03359+1523 & 0.522  & $ 7.17 \pm 1.79 $ & $ 5.28 \pm 1.32 $ & & $ 2.84 \pm 0.74 $ & $ 2.09 \pm 0.54 $ & & $ 11.0 \pm 2.56 $ & $ 8.07 \pm 1.89 $ \\
NGC2623         & 0.596  & $ 6.18 \pm 1.58 $ & $ 2.52 \pm 0.64 $ & & $ 2.38 \pm 0.64 $ & $ 0.97 \pm 0.26 $ & & $ 9.64 \pm 2.27 $ & $ 3.93 \pm 0.92 $ & 0.9$\times$0.75\tablenotemark{g,4}, 1.34\tablenotemark{j,3}\\
IRASF10173+0828 & 0.648  & $ 5.53 \pm 1.88 $ & $ 6.00 \pm 2.04 $ & & $ 2.09 \pm 0.75 $ & $ 2.27 \pm 0.82 $ & & $ 8.73 \pm 2.76 $ & $ 9.49 \pm 3.00 $ & $<3.6$\tablenotemark{c,5}\\
IC2810W         & 0.383  & $ 9.51 \pm 1.84 $ & $ 7.24 \pm 1.40 $ & & $ 3.96 \pm 0.80 $ & $ 3.02 \pm 0.61 $ & & $ 13.9 \pm 2.52 $ & $ 10.5 \pm 1.91 $ \\
NGC4194         &$<0.776$& $<4.03$           & $<0.84$           & & $<1.46          $ & $<0.30$           & & $<6.55$           & $<1.37$           & 2.1$\times$1.25\tablenotemark{h,6}, 7.7\tablenotemark{j,3}\\
NGC4418         & 0.721  & $ 4.67 \pm 2.00 $ & $ 0.83 \pm 0.35 $ & & $ 1.73 \pm 0.82 $ & $ 0.31 \pm 0.15 $ & & $ 7.51 \pm 2.95 $ & $ 1.33 \pm 0.52 $ & 0.35\tablenotemark{i,7}\\
UGC08387        &$<0.87$ & $<2.89$           & $<1.54$           & & $<1.01$           & $<0.54$           & & $<4.80$           & $<2.56$           & 2.54\tablenotemark{j,8}, 1.36\tablenotemark{f,9}\\
Mrk273          & 0.745  & $ 4.39 \pm 2.05 $ & $ 3.69 \pm 1.72 $ & & $ 1.61 \pm 0.85 $ & $ 1.35 \pm 0.71 $ & & $ 7.11 \pm 3.04 $ & $ 5.96 \pm 2.55 $ & 3.45$\times$1.7\tablenotemark{e,10}, $<1.1$\tablenotemark{e,11}, 0.58\tablenotemark{f,9}\\
IRASF14348-1447 & 0.629  & $ 5.76 \pm 1.85 $ & $ 10.8 \pm 3.47 $ & & $ 2.19 \pm 0.75 $ & $ 4.11 \pm 1.40 $ & & $ 9.06 \pm 2.72 $ & $ 17.0 \pm 5.1 $ & $1.4\times0.95$\tablenotemark{k,5}\\
CGCG049-057     &$<0.962$& $<1.50$           & $<0.48$           & & $<0.52$           & $<0.16$           & & $<2.53$           & $<0.80$           & $<1.8$\tablenotemark{c,5} \\
Arp220          & 0.630  & $ 5.74 \pm 1.86 $ & $ 2.45 \pm 0.79 $ & & $ 2.18 \pm 0.75 $ & $ 0.93 \pm 0.32 $ & & $ 9.04 \pm 2.72 $ & $ 3.85 \pm 1.16 $ & 0.7$\times$0.95\tablenotemark{b,11}, 3.5$\times$2.5\tablenotemark{b,10}, 1.4\tablenotemark{f,9}\\
IRASF17207-0014 &$>0.221$& $>14.1$           & $>13.5$           & & $>6.32$           & $>6.07$           & & $>19.1$           & $>18.3$           & $<1.0$\tablenotemark{c,5}, 0.7\tablenotemark{f,12} \\
MCG+04-48-002S  & 0.198  & $ 15.1 \pm 3.08 $ & $ 4.70 \pm 0.96 $ & & $ 6.86 \pm 1.33 $ & $ 2.14 \pm 0.41 $ & & $ 20.2 \pm 4.49 $ & $ 6.29 \pm 1.4 $ \\
IC5298          & 0.759  & $ 4.23 \pm 2.09 $ & $ 2.44 \pm 1.21 $ & & $ 1.54 \pm 0.88 $ & $ 0.89 \pm 0.50 $ & & $ 6.86 \pm 3.09 $ & $ 3.96 \pm 1.78 $ \\
NGC7771S1*      & 0.850  & $ 3.15 \pm 2.16 $ & $ 0.94 \pm 0.64 $ & & $ 1.12 \pm 1.25: $ & $ 0.33 \pm 0.37: $ & & $ 5.22 \pm 3.08 $ & $ 1.55 \pm 0.91 $ \\
Mrk331*         & 0.130  & $ 19.4 \pm 3.7  $ & $ 7.46 \pm 1.40 $ & & $ 9.20 \pm 1.63 $ & $ 3.54 \pm 0.63 $ & & $ 25.0 \pm 5.48 $ & $ 9.61 \pm 2.11 $ \\
\hline \\
Average  &  0.627 &  6.59  &  3.78  & &  2.69  &  1.51  & &  9.81  &  5.73  \\
Median	 &  0.639 &  5.66  &  2.63  & &  2.14  &  1.00  & &  8.89  &  4.12  \\
Minimum	 &  0.130 &  2.34  &  0.83  & &  0.82  &  0.31  & &  3.91  &  1.33  \\
Maxmum	 &  0.912 &  19.4  &  10.8  & &  9.21  &  4.11  & &  25.0  &  17.0  \\
\enddata
\tablecomments{
The CO radius $\mu$ is estimated from flux ratio $R_{\rm CO}$ between two observations with different telescopes.
The galaxies with $R_{\rm CO}<1$ are listed.
If $R_{\rm CO}-\sigma < 1$ for those with $R_{\rm CO} \geq 1$, 
their $R_{\rm CO}-\sigma < 1$ are listed and are used to estimate the upper limit of $\mu$.
The columns are (1): the galaxy name.
(2): the flux ratio between our observation and the literature.
(3), (4): the CO size $\mu_A$ and $Q_A$ in the Gaussian model (Model~A).
(5), (6): the CO size $\mu_B$ and $Q_B$ in the azimuthally symmetric exponential model (Model~B).
(7), (8): the CO size $\mu_C$ and $Q_C$ in the uniform disk model (Model~C).
(9): the CO radius measured with interferometers. The alphabetical superscript represents the literature:
a = \citet{Scoville1989}, b = \citet{Scoville1991}, c = \citet{Planesas1991}, d = \citet{Yun1994}, e = \citet{Yun1995}, f = \citet{DownesSolomon98},
g = \citet{BryantScoville99}, h = \citet{Aalto2000}, i = \citet{Costagliola2013}, j = \citet{Ueda14}, k = \citet{Evans00}.
The notes for each interferometric radius is given as the superscript number:
1 = deconvolved source size, double sources, 2 = deconvolved source size, bar structure, 3 = CO(2-1), radius enclosing 80\,\% of the total flux, 
4 = deconvolved core size, 5 = unresolved source, HWHM, 6 = extended component enclosing 33\,\% of the total flux, HWHM, 
7 = CO(2-1), HWHM on circle-average Gaussian in the visibility, 8 = CO(3-2), radius enclosing 80\,\% of the total flux, 
9 = CO(2-1), nuclear disk component, HWHM in Gaussian model fit, 10 = extended component, deconvolved radius, 
11 = core component, deconvolved radius, 12 = HWHM in Gaussian model fit.
The symbols of `$<$' and `$>$' indicate the upper or lower limit, respectively.
The colon represents the large error more than 100\,\%.
The asterisk of NGC~7771S1 and Mrk~331 means the model uncertainty and the uncertain CO fluxes, respectively. 
See the note in Section \ref{apx:note_obj}.
The statistics is also shown in the lower part except those with the upper/lower limit value.
}
\end{deluxetable}
\end{turnpage}

\clearpage
\begin{deluxetable*}{lllrrr}
\tabletypesize{\scriptsize}
\tablewidth{0pt}
\tablecaption{The CO size measured with the mapping observations\label{tab:mapping}}
\tablehead{
\colhead{Galaxies} & \colhead{R.A.(J2000)} & \colhead{Dec.(J2000)} & \colhead{$D_{\rm L}$} & \colhead{Deconvolved HWHM} & \colhead{Deconvolved HWHM} \\ 
\colhead{} & \colhead{} & \colhead{} & \colhead{(Mpc)} & \colhead{(arcsec)} & \colhead{(kpc)} \\
\colhead{(1)} & \colhead{(2)} & \colhead{(3)} & \colhead{(4)} & \colhead{(5)} & \colhead{(6)}
} 
\startdata
CGCG011-076 & $11^{\rm h}21^{\rm m}12.24^{\rm s}$ & $-02^{\circ}59^{\prime}01.9^{\prime\prime}$ & 117 & 10.6 & 6.03  \\
CGCG052-037 & $16^{\rm h}30^{\rm m}56.60^{\rm s}$ & $+04^{\circ}04^{\prime}58.4^{\prime\prime}$ & 116 & 10.2 & 5.74  \\
MCG-02-33-098 & $13^{\rm h}02^{\rm m}19.80^{\rm s}$ & $-15^{\circ}46^{\prime}03.5^{\prime\prime}$ & 78.7 & $<6.77$ & $<2.58$  \\
NGC5653 & $14^{\rm h}30^{\rm m}09.89^{\rm s}$ & $+31^{\circ}12^{\prime}56.33^{\prime\prime}$ & 60.2 & 24.1 & 7.03  \\
NGC5990 & $15^{\rm h}46^{\rm m}16.40^{\rm s}$ & $+02^{\circ}24^{\prime}54.70^{\prime\prime}$ & 64.4 & 3.33 & 1.04  \\
IC0563N & $09^{\rm h}46^{\rm m}21.06^{\rm s}$ & $+03^{\circ}04^{\prime}16.6^{\prime\prime}$ & 92.9 & 8.69 & 3.91 \\
UGC03351 & $05^{\rm h}45^{\rm m}48.01^{\rm s}$ & $+58^{\circ}42^{\prime}03.7^{\prime\prime}$ & 65.8 & 14.9 & 4.74 \\
\enddata
\tablecomments{
Col. (1): Name of galaxies whose CO size are measured from the mapping observations. 
Col. (2), (3): The source coordinates.
Col. (4): The luminosity distance taken from \citet{Armus09}.
Col. (5), (6): Deconvolved HWHMs in arcsec and kpc scale. See Appendix \ref{apx:mapping} for details.
The symbol of `$<$' indicates the upper limit.
}
\end{deluxetable*}


\begin{thebibliography}{}
\bibitem[Aalto \& H{\"u}ttemeister(2000)]{Aalto2000} Aalto, S., \& H{\"u}ttemeister, S.\ 2000, \aap, 362, 42 
\bibitem[Armus et al.(2007)]{Armus07} Armus, L., Charmandaris, V., Bernard-Salas, J., et al.\ 2007, \apj, 656, 148 
\bibitem[Armus et al.(2009)]{Armus09} Armus, L., Mazzarella, J.~M., Evans, A.~S., et al.\ 2009, \pasp, 121, 559 
\bibitem[Barnes \& Hernquist(1996)]{BarnesHernquist96} Barnes, J.~E., \& Hernquist, L.\ 1996, \apj, 471, 115 
\bibitem[Bolatto et al.(2013)]{Bolatto13} Bolatto, A.~D., Wolfire, M., \& Leroy, A.~K.\ 2013, \araa, 51, 207 
\bibitem[Bryant \& Scoville(1999)]{BryantScoville99} Bryant, P.~M., \& Scoville, N.~Z.\ 1999, \aj, 117, 2632
\bibitem[Caputi et al.(2007)]{Caputi07} Caputi, K.~I., Lagache, G., Yan, L., et al.\ 2007, \apj, 660, 97 
\bibitem[Casey(2012)]{Casey12} Casey, C.~M.\ 2012, \mnras, 425, 3094 
\bibitem[Cazzoli et al.(2014)]{Cazzoli14} Cazzoli, S., Arribas, S., Colina, L., et al.\ 2014, \aap, 569, A14 
\bibitem[Cicone et al.(2014)]{Cicone14} Cicone, C., Maiolino, R., Sturm, E., et al.\ 2014, \aap, 562, A21 
\bibitem[Costagliola et al.(2013)]{Costagliola2013} Costagliola, F., Aalto, S., Sakamoto, K., et al.\ 2013, \aap, 556, A66 
\bibitem[Cox et al.(2008)]{Cox08} Cox, T.~J., Jonsson, P., Somerville, R.~S., Primack, J.~R., \& Dekel, A.\ 2008, \mnras, 384, 386 
\bibitem[Curran et al.(2000)]{Curran00} Curran, S.~J., Aalto, S., \& Booth, R.~S.\ 2000, \aaps, 141, 193 
\bibitem[Daddi et al.(2010)]{Daddi10b} Daddi, E., Elbaz, D., Walter, F., et al.\ 2010, \apjl, 714, L118 
\bibitem[Dale et al.(2005)]{Dale05} Dale, D.~A., Sheth, K., Helou, G., Regan, M.~W., H{\"u}ttemeister, S.\ 2005, \aj, 129, 2197 
\bibitem[D{\'{\i}}az-Santos et al.(2010)]{Diaz-Santos10} D{\'{\i}}az-Santos, T., Charmandaris, V., Armus, L., et al.\ 2010, \apj, 723, 993 
\bibitem[D{\'{\i}}az-Santos et al.(2011)]{Diaz-Santos11} D{\'{\i}}az-Santos, T., Charmandaris, V., Armus, L., et al.\ 2011, \apj, 741, 32 
\bibitem[Dinh-V-Trung et al.(2001)]{Dinh2001} Dinh-V-Trung,, Lo, K.~Y., Kim, D.-C., Gao, Y., \& Gruendl, R.~A.\ 2001, \apj, 556, 141 
\bibitem[Downes et al.(1993)]{Downes93} Downes, D., Solomon, P.~M., \& Radford, S.~J.~E.\ 1993, \apjl, 414, L13 
\bibitem[Downes \& Solomon(1998)]{DownesSolomon98} Downes, D., \& Solomon, P.~M.\ 1998, \apj, 507, 615 
\bibitem[Elbaz et al.(2002)]{Elbaz02} Elbaz, D., Cesarsky, C.~J., Chanial, P., et al.\ 2002, \aap, 384, 848 
\bibitem[Elfhag et al.(1996)]{Elfhag96} Elfhag, T., Booth, R.~S., Hoeglund, B., Johansson, L.~E.~B., \& Sandqvist, A.\ 1996, \aaps, 115, 439 
\bibitem[Evans et al.(2000)]{Evans00} Evans, A.~S., Surace, J.~A., \& Mazzarella, J.~M.\ 2000, \apjl, 529, L85 
\bibitem[Evans et al.(2005)]{Evans05} Evans, A.~S., Mazzarella, J.~M., Surace, J.~A., et al.\ 2005, \apjs, 159, 197 
\bibitem[Gao et al.(1997a)]{Gao97BIMA} Gao, Y., Gruendl, R., Lo, K.~Y., Hwang, C.~Y., \& Veilleux, S.\ 1997, American Institute of Physics Conference Series, 393, 319
\bibitem[Gao et al.(1997b)]{Gao97} Gao, Y., Solomon, P.~M., Downes, D., \& Radford, S.~J.~E.\ 1997, \apjl, 481, L35 
\bibitem[Gao \& Solomon(1999)]{GaoSolomon99} Gao, Y., \& Solomon, P.~M.\ 1999, \apjl, 512, L99 
\bibitem[Gao et al.(2001)]{Gao01} Gao, Y., Lo, K.~Y., Lee, S.-W., \& Lee, T.-H.\ 2001, \apj, 548, 172 
\bibitem[Gao \& Solomon(2004)]{GaoSolomon04a} Gao, Y., \& Solomon, P.~M.\ 2004, \apjs, 152, 63 
\bibitem[Garc{\'{\i}}a-Burillo et al.(2014)]{Garcia14} Garc{\'{\i}}a-Burillo, S., Combes, F., Usero, A., et al.\ 2014, \aap, 567, A125 
\bibitem[Garc{\'{\i}}a-Burillo et al.(2012)]{Garcia12} Garc{\'{\i}}a-Burillo, S., Usero, A., Alonso-Herrero, A., et al.\ 2012, \aap, 539, A8 
\bibitem[Genzel et al.(2012)]{Genzel12} Genzel, R., Tacconi, L.~J., Combes, F., et al.\ 2012, \apj, 746, 69 
\bibitem[Glover \& Mac Low(2011)]{GloverMacLow11} Glover, S.~C.~O., \& Mac Low, M.-M.\ 2011, \mnras, 412, 337 
\bibitem[Goto et al.(2010)]{Goto10} Goto, T., Takagi, T., Matsuhara, H., et al.\ 2010, \aap, 514, A6 
\bibitem[Haan et al.(2011)]{Haan11} Haan, S., Surace, J.~A., Armus, L., et al.\ 2011, \aj, 141, 100
\bibitem[Howell et al.(2010)]{Howell10} Howell, J.~H., Armus, L., Mazzarella, J.~M., et al.\ 2010, \apj, 715, 572 
\bibitem[Imanishi \& Nakanishi(2013)]{ImanishiNkanishi13} Imanishi, M., \& Nakanishi, K.\ 2013, \aj, 146, 47
\bibitem[Inami et al.(2013)]{Inami13} Inami, H., Armus, L., Charmandaris, V., et al.\ 2013, \apj, 777, 156 
\bibitem[Iono et al.(2005)]{Iono05} Iono, D., Yun, M.~S., \& Ho, P.~T.~P.\ 2005, \apjs, 158, 1 
\bibitem[Iono et al.(2004)]{Iono04} Iono, D., Yun, M.~S., \& Mihos, J.~C.\ 2004, \apj, 616, 199 
\bibitem[Kamazaki et al.(2012)]{Kamazaki12} Kamazaki, T., Okumura, S.~K., Chikada, Y., et al.\ 2012, \pasj, 64, 29 
\bibitem[Kapferer et al.(2005)]{Kapferer05} Kapferer, W., Knapp, A., Schindler, S., Kimeswenger, S., \& van Kampen, E.\ 2005, \aap, 438, 87 
\bibitem[Kennicutt(1998)]{Kennicutt98} Kennicutt, R.~C., Jr.\ 1998, \araa, 36, 189 
\bibitem[Kim et al.(1998)]{Kim98} Kim, D.-C., Veilleux, S., \& Sanders, D.~B.\ 1998, \apj, 508, 627
\bibitem[Lazareff et al.(1989)]{Lazareff89} Lazareff, B., Castets, A., Kim, D.-W., \& Jura, M.\ 1989, \apjl, 336, L13 
\bibitem[Le Floc'h et al.(2005)]{LeFloch05} Le Floc'h, E., Papovich, C., Dole, H., et al.\ 2005, \apj, 632, 169 
\bibitem[Leech et al.(2010)]{Leech10} Leech, J., Isaak, K.~G., Papadopoulos, P.~P., Gao, Y., \& Davis, G.~R.\ 2010, \mnras, 406, 1364 
\bibitem[Lo et al.(1997)]{Lo97} Lo, K.~Y., Gao, Y., \& Gruendl, R.~A.\ 1997, \apjl, 475, L103 
\bibitem[Magnelli et al.(2009)]{Magnelli09} Magnelli, B., Elbaz, D., Chary, R.~R., et al.\ 2009, \aap, 496, 57 
\bibitem[Magnelli et al.(2013)]{Magnelli13} Magnelli, B., Popesso, P., Berta, S., et al.\ 2013, \aap, 553, AA132 
\bibitem[Magnelli et al.(2012)]{Magnelli12} Magnelli, B., Saintonge, A., Lutz, D., et al.\ 2012, \aap, 548, A22 
\bibitem[Mazzarella et al.(1993)]{Mazzarella93} Mazzarella, J.~M., Graham, J.~R., Sanders, D.~B., \& Djorgovski, S.\ 1993, \apj, 409, 170 
\bibitem[Mihos \& Hernquist(1996)]{MihosHernquist96} Mihos, J.~C., \& Hernquist, L.\ 1996, \apj, 464, 641
\bibitem[Mirabel et al.(1990)]{Mirabel90} Mirabel, I.~F., Booth, R.~S., Johansson, L.~E.~B., Garay, G., \& Sanders, D.~B.\ 1990, \aap, 236, 327 
\bibitem[Mirabel et al.(1989)]{Mirabel89} Mirabel, I.~F., Sanders, D.~B., \& Kazes, I.\ 1989, \apjl, 340, L9 
\bibitem[Nakajima et al.(2013)]{Nakajima13} Nakajima, T., Kimura, K., Nishimura, A., et al.\ 2013, \pasp, 125, 252 
\bibitem[Nakajima et al.(2008)]{Nakajima08} Nakajima, T., Sakai, T., Asayama, S., et al.\ 2008, \pasj, 60, 435 
\bibitem[Nakanishi (2005)]{Nakanishi2005} Nakanishi, H. 2005, PhD Thesis, The University of Tokyo
\bibitem[Nakanishi \& Sofue(2006)]{NakanishiSofue2006} Nakanishi, H., \& Sofue, Y.\ 2006, \pasj, 58, 847 
\bibitem[Nishiyama et al.(2001)]{Nishiyama01b} Nishiyama, K., Nakai, N., \& Kuno, N.\ 2001, \pasj, 53, 757 
\bibitem[Papadopoulos et al.(2012)]{Papadopoulos12} Papadopoulos, P.~P., van der Werf, P.~P., Xilouris, E.~M., et al.\ 2012, \mnras, 426, 2601 
\bibitem[Petric et al.(2011)]{Petric11} Petric, A.~O., Armus, L., Howell, J., et al.\ 2011, \apj, 730, 28 
\bibitem[Planesas et al.(1991)]{Planesas1991} Planesas, P., Mirabel, I.~F., \& Sanders, D.~B.\ 1991, \apj, 370, 172 
\bibitem[Saintonge et al.(2011)]{Saintonge11} Saintonge, A., Kauffmann, G., Kramer, C., et al.\ 2011, \mnras, 415, 32 
\bibitem[Saito et al.(2015)]{Saito15} Saito, T., Iono, D., Yun, M.~S., et al.\ 2015, \apj, 803, 60 
\bibitem[Sakamoto et al.(2014)]{Sakamoto14} Sakamoto, K., Aalto, S., Combes, F., Evans, A., \& Peck, A.\ 2014, arXiv:1403.7117 
\bibitem[Sakamoto et al.(1999)]{Sakamoto99} Sakamoto, K., Okumura, S.~K., Ishizuki, S., \& Scoville, N.~Z.\ 1999, \apj, 525, 691 
\bibitem[Salom{\'e} et al.(2006)]{Salome06} Salom{\'e}, P., Combes, F., Edge, A.~C., et al.\ 2006, \aap, 454, 437 
\bibitem[Salom{\'e} et al.(2008b)]{Salome08b} Salom{\'e}, P., Combes, F., Revaz, Y., et al.\ 2008, \aap, 484, 317 
\bibitem[Salom{\'e} et al.(2008a)]{Salome08a} Salom{\'e}, P., Revaz, Y., Combes, F., et al.\ 2008, \aap, 483, 793 
\bibitem[Sanders et al.(2003)]{Sanders03} Sanders, D.~B., Mazzarella, J.~M., Kim, D.-C., Surace, J.~A., \& Soifer, B.~T.\ 2003, \aj, 126, 1607
\bibitem[Sanders \& Mirabel(1985)]{SandersMirabel85} Sanders, D.~B., \& Mirabel, I.~F.\ 1985, \apjl, 298, L31 
\bibitem[Sanders \& Mirabel(1996)]{SandersMirabel96} Sanders, D.~B., \& Mirabel, I.~F.\ 1996, \araa, 34, 749 
\bibitem[Sanders et al.(1991)]{Sanders91} Sanders, D.~B., Scoville, N.~Z., \& Soifer, B.~T.\ 1991, \apj, 370, 158 
\bibitem[Sanders et al.(1986)]{Sanders86} Sanders, D.~B., Scoville, N.~Z., Young, J.~S., et al.\ 1986, \apjl, 305, L45 
\bibitem[Sanders et al.(1984)]{Sanders84} Sanders, D.~B., Solomon, P.~M., \& Scoville, N.~Z.\ 1984, \apj, 276, 182 
\bibitem[Sandstrom et al.(2013)]{Sandstrom13} Sandstrom, K.~M., Leroy, A.~K., Walter, F., et al.\ 2013, \apj, 777, 5 
\bibitem[Scoville et al.(1989)]{Scoville1989} Scoville, N.~Z., Sanders, D.~B., Sargent, A.~I., Soifer, B.~T., \& Tinney, C.~G.\ 1989, \apjl, 345, L25 
\bibitem[Scoville et al.(1991)]{Scoville1991} Scoville, N.~Z., Sargent, A.~I., Sanders, D.~B., \& Soifer, B.~T.\ 1991, \apjl, 366, L5 
\bibitem[Scoville et al.(2014)]{Scoville14arxiv} Scoville, N., Sheth, K., Walter, F., et al.\ 2014, arXiv:1412.5183 
\bibitem[Skrutskie et al.(2006)]{Skrutskie06} Skrutskie, M.~F., Cutri, R.~M., Stiening, R., et al.\ 2006, \aj, 131, 1163 
\bibitem[Smith et al.(1999)]{Smith99} Smith, D.~A., Herter, T., Haynes, M.~P., \& Neff, S.~G.\ 1999, \apj, 510, 669 
\bibitem[Solomon et al.(1997)]{Solomon97} Solomon, P.~M., Downes, D., Radford, S.~J.~E., \& Barrett, J.~W.\ 1997, \apj, 478, 144 
\bibitem[Stierwalt et al.(2014)]{Stierwalt14} Stierwalt, S., Armus, L., Charmandaris, V., et al.\ 2014, \apj, 790, 124 
\bibitem[Stierwalt et al.(2013)]{Stierwalt13} Stierwalt, S., Armus, L., Surace, J.~A., et al.\ 2013, \apjs, 206, 1
\bibitem[Tacconi et al.(2008)]{Tacconi08} Tacconi, L.~J., Genzel, R., Smail, I., et al.\ 2008, \apj, 680, 246 
\bibitem[U et al.(2012)]{U12} U, V., Sanders, D.~B., Mazzarella, J.~M., et al.\ 2012, \apjs, 203, 9 
\bibitem[Ueda et al.(2014)]{Ueda14} Ueda, J., Iono, D., Yun, M.~S., et al.\ 2014, \apjs, 214, 1 
\bibitem[Walter et al.(2002)]{Walter02} Walter, F., Weiss, A., \& Scoville, N.\ 2002, \apjl, 580, L21 
\bibitem[Wang et al.(2001)]{Wang01} Wang, W.-H., Lo, K.~Y., Gao, Y., \& Gruendl, R.~A.\ 2001, \aj, 122, 140 
\bibitem[Wolfire et al.(2010)]{Wolfire10} Wolfire, M.~G., Hollenbach, D., \& McKee, C.~F.\ 2010, \apj, 716, 1191 
\bibitem[Yao et al.(2003)]{Yao03} Yao, L., Seaquist, E.~R., Kuno, N., \& Dunne, L.\ 2003, \apj, 588, 771 
\bibitem[Young et al.(1986)]{Young86} Young, J.~S., Schloerb, F.~P., Kenney, J.~D., \& Lord, S.~D.\ 1986, \apj, 304, 443 
\bibitem[Young \& Scoville(1982)]{YoungScoville1982} Young, J.~S., \& Scoville, N.\ 1982, \apj, 258, 467 
\bibitem[Young et al.(1989)]{Young89} Young, J.~S., Xie, S., Kenney, J.~D.~P., \& Rice, W.~L.\ 1989, \apjs, 70, 699 
\bibitem[Young et al.(1995)]{Young95} Young, J.~S., Xie, S., Tacconi, L., et al.\ 1995, \apjs, 98, 219 
\bibitem[Yun et al.(1994)]{Yun1994} Yun, M.~S., Scoville, N.~Z., \& Knop, R.~A.\ 1994, \apjl, 430, L109 
\bibitem[Yun \& Scoville(1995)]{Yun1995} Yun, M.~S., \& Scoville, N.~Z.\ 1995, \apjl, 451, L45 
\bibitem[Xu et al.(2014)]{Xu14} Xu, C.~K., Cao, C., Lu, N., et al.\ 2014, \apj, 787, 48 
\bibitem[Xu et al.(2015)]{Xu15} Xu, C.~K., Cao, C., Lu, N., et al.\ 2015, \apj, 799, 11 
\end{thebibliography}
\end{document}